\let\latexaddtocontents\addtocontents
\documentclass[a4paper,twocolumn,11pt,unpublished=2025-07-11, accepted=2026-03-10]{quantumarticle}
\let\addtocontents\latexaddtocontents
\pdfoutput=1
\usepackage[utf8]{inputenc}
\usepackage[english]{babel}
\usepackage[T1]{fontenc}

\usepackage[backend=bibtex, sorting=none]{biblatex}
\bibliography{references}
\usepackage{amsmath}
\usepackage{amssymb}
\usepackage{hyperref}

\usepackage{tikz}
\usepackage{lipsum}
\usepackage{mathrsfs}
\usepackage{float}
\usepackage{soul}
\usepackage{physics}
\usepackage{dsfont}

\usepackage{pgfplots}
\usepackage{newunicodechar}
\usetikzlibrary{decorations.pathmorphing}
\definecolor{electricviolet}{rgb}{0.56, 0.0, 1.0}
\definecolor{amaranth}{rgb}{0.9, 0.17, 0.31}
\newcommand{\T}{\tau}

\def\be{\begin{equation}}
\def\ee{\end{equation}}
\pgfmathdeclarefunction{gauss}{2}{%
  \pgfmathparse{1/(#2*sqrt(2*pi))*exp(-((x-#1)^2)/(2*#2^2))}%
}
\pgfplotsset{compat=1.18}
\begin{document}

\title{CDJ-Pontryagin Optimal Control for General Continuously Monitored Quantum Systems}

\author{Tathagata Karmakar}
	\email{tatha@berkeley.edu}
	\affiliation{Department of Physics and Astronomy, University of Rochester, Rochester, NY 14627, USA}
	\affiliation{Center for Coherence and Quantum Optics, University of Rochester, Rochester, NY 14627, USA}
 
 \affiliation{Institute for Quantum Studies, Chapman University, Orange, CA, 92866, USA}
 \affiliation{Department of Chemistry, University of California, Berkeley, California 94720, USA}
 \affiliation{Berkeley Center for Quantum Information and Computation, Berkeley, California 94720, USA}
	\author{Andrew N. Jordan}
 \email{jordan@chapman.edu}
 \affiliation{Institute for Quantum Studies, Chapman University, Orange, CA, 92866, USA}
	\affiliation{Department of Physics and Astronomy, University of Rochester, Rochester, NY 14627, USA}
	\affiliation{Center for Coherence and Quantum Optics, University of Rochester, Rochester, NY 14627, USA}  
    \affiliation{The Kennedy Chair in Physics, Chapman University, Orange, CA 92866, USA}
\maketitle

\begin{abstract}
 The Chantasri-Dressel-Jordan (CDJ) stochastic path integral formalism  (Chantasri et al.~2013 and 2015)   characterizes the statistics of the readouts and the most likely conditional evolution of continuously monitored quantum systems involving a few qubits or quantum harmonic oscillators in Gaussian states.  In our work, we generalize the CDJ formalism to arbitrary continuously monitored systems by introducing a costate operator. We then prescribe a generalized Pontryagin's maximum principle for quantum systems undergoing arbitrary evolution and find conditions on optimal control protocols. We show that the CDJ formalism's most likely path can be cast as a quantum Pontryagin's maximum principle, where the cost function is the readout probabilities along a quantum trajectory.  This insight allows us to derive general optimal control equations for arbitrary control parameters to achieve a given task.  We apply our results to a monitored oscillator in the presence of a parametric quadratic potential and variable quadrature measurements. We find the optimal potential strength and quadrature angle for fixed-end point problems. The optimal parametric potential is analytically shown to have a ``bang-bang'' form. We apply our protocol to three quantum oscillator examples relevant to Bosonic quantum computing.   The first example considers a binomial codeword preparation from an error word, the second example looks into cooling to the ground state from an even cat state, and the third example investigates a cat state to cat state evolution. We compare the statistics of the fidelities of the final state with respect to the target state for trajectories generated under the optimal control with those generated under a sample control. Compared to the latter case, we see a 40-196\%  increase in the number of trajectories reaching more than 95\% fidelities under the optimal control. Our work provides a systematic prescription for finding quantum optimal control for continuously monitored systems. 
\end{abstract}
\section{\label{sec:level1}Introduction}
Near-term quantum hardware needs high-fidelity state preparation or stabilization schemes \cite{Preskill2018quantumcomputingin}. State preparation schemes are required for tasks such as encoding error correcting codes \cite{10.21468/SciPostPhysLectNotes.70}, running quantum algorithms,  performing simulations \cite{RevModPhys.90.015002, RevModPhys.94.015004}, etc.    To that end, a promising approach is dissipation engineering.  The idea is to utilize a carefully designed system-environment interaction as a resource for quantum computation  \cite{verstraete_quantum_2009,harrington_engineered_2022}. Experiments have utilized dissipation engineering to prepare quantum states \cite{PhysRevLett.121.060502, PRXQuantum.5.020321}, stabilize quantum states \cite{doi:10.1126/science.aaa2085, li_autonomous_2024}, produce entangled states \cite{lin_dissipative_2013,PhysRevLett.128.080502}, perform autonomous quantum error correction \cite{gertler_protecting_2021, li_aqec}, etc. However, finding optimal control for such tasks can be challenging \cite{PhysRevLett.132.040404}.

In this article, we investigate optimal control protocols for quantum systems undergoing measurements. Quantum measurements are performed by monitoring the environment coupled to a quantum system \cite{BookWiseman, BookJordan}. When the system-environment coupling is weak, and the environment is measured in small successive time intervals, quasi-continuous monitoring of the quantum system is realized. In this scenario, the system's evolution conditioned on the measurement readouts is stochastic. The sequence of readouts and associated conditional state evolution realizes a quantum trajectory \cite{BookJordan,brun_simple_2002, jacobs2006straightforward}. Continuous measurements allow real-time tracking of a quantum state \cite{PhysRevA.78.062322, PhysRevA.102.062418, PhysRevA.102.062219, PhysRevLett.96.010504}.  
 Such measurements have been utilized in a diverse set of scenarios, such as qubit state monitoring \cite{murch2013observing, PhysRevLett.114.090403},  remote entanglement generation \cite{PhysRevLett.115.150503, PhysRevLett.112.170501}, observation of quantum state diffusion from a fluorescing
qubit \cite{PhysRevX.6.011002},  Zeno dragging of a qubit \cite{PhysRevLett.120.020505}, monitoring of trapped atoms \cite{colangelo_simultaneous_2017}, quantum sensing \cite{doi:10.1126/science.aam7009, PhysRevApplied.14.014013}, catching and reversing a quantum jump \cite{minev_catch_2019}, continuous error correction \cite{livingston_experimental_2022}, etc.    In our work, we consider quantum optimal control for continuously monitored systems. For simplicity, the control parameters in our analysis correspond to either a tunable system-environment interaction (i.e., quadrature of a homodyne readout) or the strength of a drive applied to the system.  To find the optimal control protocols, we generalize the Chantasri-Dressel-Jordan (CDJ) most likely path-based approach previously adopted in Refs.~\cite{kokaew_quantum_2026, lewalle2023optimal}.
 
The CDJ formalism expresses the probability densities of the quantum trajectories of a continuously monitored system in terms of a path integral of the exponential of a stochastic action  \cite{PhysRevA.88.042110,PhysRevA.92.032125,Areeya_Thesis}. The stochastic action is a functional of the readouts, a set of coordinates characterizing the quantum state (e.g.~Bloch coordinates for a qubit), and a corresponding set of auxiliary variables.  The trajectories corresponding to the most likely set of readouts are called most likely paths (MLP) or optimal paths (OP). We can express the most likely evolution in terms of Hamilton's equations between state coordinates and the auxiliary variables. Thus, the CDJ formalism provides a classical-like description of continuously monitored systems. The CDJ stochastic path integral approach has led to insights into the statistics of the quantum trajectories of monitored qubits \cite{jordan_anatomy_2016, FlorTeach2019, LewalleMultipath, LewalleChaos, shea2023action, shea2024stochastic}.   The predictions of the CDJ stochastic path integral formulation have been verified experimentally as well \cite{weber_mapping_2014, NaghilooCaustic, PhysRevX.6.041052}. Previously, we extended the CDJ stochastic path integral approach to continuous variable systems \cite{PRXQuantum.3.010327}. Namely, to a simple harmonic oscillator in general Gaussian states undergoing simultaneous continuous position and momentum measurements. Under the steady-state assumption, we analytically solved for the optimal paths and confirmed our results with simulated trajectories. Recent investigations indicate that the CDJ formalism can provide an effective strategy for finding optimal controls. Kokaew et al.~adopted the CDJ description to find most likely path-based optimal control protocols for state preparation \cite{kokaew_quantum_2026}. Their investigations suggested that state preparation using the most likely paths for noisy quantum systems can lead to higher success rates than the Lindbladian master equation. Lewalle et al.~have looked into Pontryagin maximum principle (PMP) based optimal control for qubits undergoing non-Hermitian dynamics \cite{PhysRevA.107.022216}.  The Pontryagin maximum principle provides a strategy for optimal control under constrained evolution by introducing costate parameters \cite{PRXQuantum.2.030203}. Lewalle et al.~also provided a connection between CDJ stochastic path integral formulation and the PMP and demonstrated the optimality of the shortcut to Zeno approaches (with counter-diabatic driving) for Zeno dragging of a qubit \cite{lewalle2023optimal}. However, such approaches have relied on differentiating the CDJ action with respect to the control parameter. For bounded controls, the optimality is often achieved at the boundary, as we will see shortly. In such cases, the applicability of the previous results is limited.

Despite its successes, the CDJ stochastic path integral approach does have drawbacks. Construction of the stochastic action uses a parametrization of the state of the quantum system in terms of a finite number of variables (e.g., Bloch coordinates of a qubit or position and momentum expectation values of an oscillator). This can be a severe limitation for many-body and continuous-variable systems in non-Gaussian states. In this work, we overcome this limitation by providing a stochastic action principle for general continuously monitored systems.   We achieve this by taking inspiration from the Pontryagin maximum principle and introducing a costate operator. Such a costate-based description was also prescribed in the Appendix of Ref.~\cite{jordan_anatomy_2016}. Here, we first reproduce our previous most likely path work for Gaussian states of a harmonic oscillator.  Then, we provide a general formulation of Pontryagin's maximum principle for optimizing an arbitrary cost functional constrained on the evolution of a quantum system under control. We derive the Chantasri-Dressel-Jordan most-likely paths as a special case of the general Pontryagin's maximum principle. Next, we examine the case of a quantum harmonic oscillator in the presence of a parametric quadratic potential and undergoing variable quadrature measurements. For this system, we analytically find the optimal parametric potential strength and measurement quadrature under fixed-endpoint constraints. The optimal parametric potential is found to be of ``bang-bang'' form. We examine three examples of state preparation problems. First, we look at the preparation of the binomial code word $\frac{\ket{0}+\ket{4}} {\sqrt{2}}$ starting from the state $\frac{\ket{0}-\ket{4}} {\sqrt{2}}$. Binomial codes represent logical qubits (in other words,  logical code words) in terms of coherent superpositions of the Fock states of an oscillator  \cite{PhysRevX.6.031006, 10.21468/SciPostPhysLectNotes.70}. Such codes can protect against single or multiple photon loss and dephasing errors, and are crucial for error correction with continuous variable quantum systems.   In the second example, we consider cooling to the ground state, starting from a cat state \cite{gerry_quantum_1997, PhysRevA.107.023516, PhysRevA.107.053521}. As our third example, we look at a cat state to cat state evolution where the sizes are $|\alpha|^2\sim 2.4$. 
The examples we analyze are relevant for tasks such as error correction, ground state preparation with oscillators \cite{rossi_measurement-based_2018,doi:10.1126/science.abh2634}. In all three cases, we construct a sample control that can also help us reach the target state. We then generate 10,000 stochastic trajectories for both the optimal and the sample control. From those simulated trajectories, we compare the histograms of the final state fidelities with respect to the target state. We find, compared to the sample control, the optimal control leads to a 40-196\% increase in the number of trajectories reaching more than 95\% fidelity. The results show that optimal controls obtained under the fixed-endpoints assumptions can be helpful in state preparation.  We stress that although we restrict the examples to a continuously monitored harmonic oscillator, the general principles presented are applicable beyond monitored quantum systems.

The utility of the costate formalism for finding quantum optimal control of open quantum systems is known \cite{Koch_2016}. Using the Pontryagin maximum principle, the optimality of quantum annealing (QA) and quantum approximate optimization algorithm (QAOA) has been investigated  \cite{PhysRevApplied.16.054023}. The costate-based approach has helped optimize transmon readout as well \cite{gautier2024optimal}. The Pontryagin maximum principle has been applied to variational quantum algorithms where the optimality of ``bang-bang'' protocols has been shown \cite{PhysRevX.7.021027, PRXQuantum.2.010101}. Other problems where the principle has been employed include maximally entangled state preparation \cite{PhysRevA.97.062343}, and  Grover's search algorithm \cite{PhysRevA.100.022327}, pulse optimization in circuit-QED systems  \cite{zhou2024optimalcontrolopenquantum}. Such a general framework for continuously monitored systems was lacking. Furthermore, past investigations considered open quantum systems with linear evolution. In such cases, the costate evolution equation was found to be the negative conjugate of the state evolution equation  \cite{Koch_2016, PhysRevApplied.16.054023}. However, the conditional evolution of a monitored quantum system is nonlinear. Thus, in our work, we obtain a more complicated and nonlinear relationship between the state and costate evolutions. We also show that when the cost function corresponds to readout probabilities,  an operator transformation can reduce the costate evolution equation to the negative conjugate of the state evolution. Such simplification might not follow for the general cost functions. 

We restrict our attention to a monitored harmonic oscillator as it provides a suitable description of a wide range of systems. These include mirrors, membranes, optical degrees of freedom, superconducting resonators, and so on. Mechanical resonators can be monitored by clamping them in a cavity and performing homodyne measurements on the transmitted light  \cite{RevModPhys.86.1391,RevModPhys.82.1155, rossi_measurement-based_2018, PhysRevLett.123.163601}. Both theoretical and experimental investigations into oscillator monitoring \cite{vanner_towards_2015,PhysRevLett.108.033602,PhysRevLett.117.140401,PhysRevX.7.021008,moller_quantum_2017,PhysRevX.7.011001,PhysRevLett.123.163601, purdy_observation_2013,thompson, jacobs2006straightforward,Genoni2016,belenchia_entropy_2020,PhysRevA.60.2700}, and feedback control \cite{rossi_measurement-based_2018,wilson_measurement-based_2015,krause2015optical,PhysRevLett.99.017201,kleckner_sub-kelvin_2006,PhysRevLett.99.160801} have been done. Explorations involving mechanical resonators include the generation of superposition states \cite{ringbauer_generation_2018}, squeezed states  \cite{lei_quantum_2016, wollman_quantum_2015,PhysRevLett.115.243601,PhysRevX.5.041037, PhysRevLett.112.023601},   resonator-resonator entanglement \cite{Riedinger:2018qsr,ockeloen-korppi_stabilized_2018}, optical field-resonator entanglement  \cite{riedinger_non-classical_2016,palomaki_entangling_2013}, etc. Superconducting cavity resonators \cite{RevModPhys.93.025005, doi:10.1073/pnas.2221736120, chou2023demonstratingsuperconductingdualrailcavity} and nonlinear optical systems \cite{Yanagimoto:22, Yanagimoto:24, PRXQuantum.4.010333, yanagimoto2023engineeringcubicquantumnondemolition, PhysRevLett.124.240503}   have witnessed significant progress as well.   Therefore, the examples discussed in our work are relevant to a broad set of ongoing investigations.

This paper is organized as follows. In section \ref{stoch_evol}, we describe the stochastic evolution of a monitored quantum system, provide a brief introduction to the Chantasri-Dressel-Jordan stochastic path integral formalism, construct the stochastic Hamiltonian for a general monitored system, and verify the optimal paths predicted by the general formalism with previous results for a monitored harmonic oscillator. Section \ref{text_pontrya} formulates Pontryagin's maximum principle for a general quantum system and rederives the general CDJ most likely paths as a special case. Section \ref{text_pontrya} also looks at the optimal control problem for a monitored oscillator, provides associated equations to be solved, and analyzes three examples.  Lastly, we conclude in section \ref{conclusions}.

\section{\label{stoch_evol}Stochastic evolution of the system}
\begin{figure}
    \centering
\begin{tikzpicture}
\node[] at (3.95,0.2) {\includegraphics[width=0.13\linewidth]{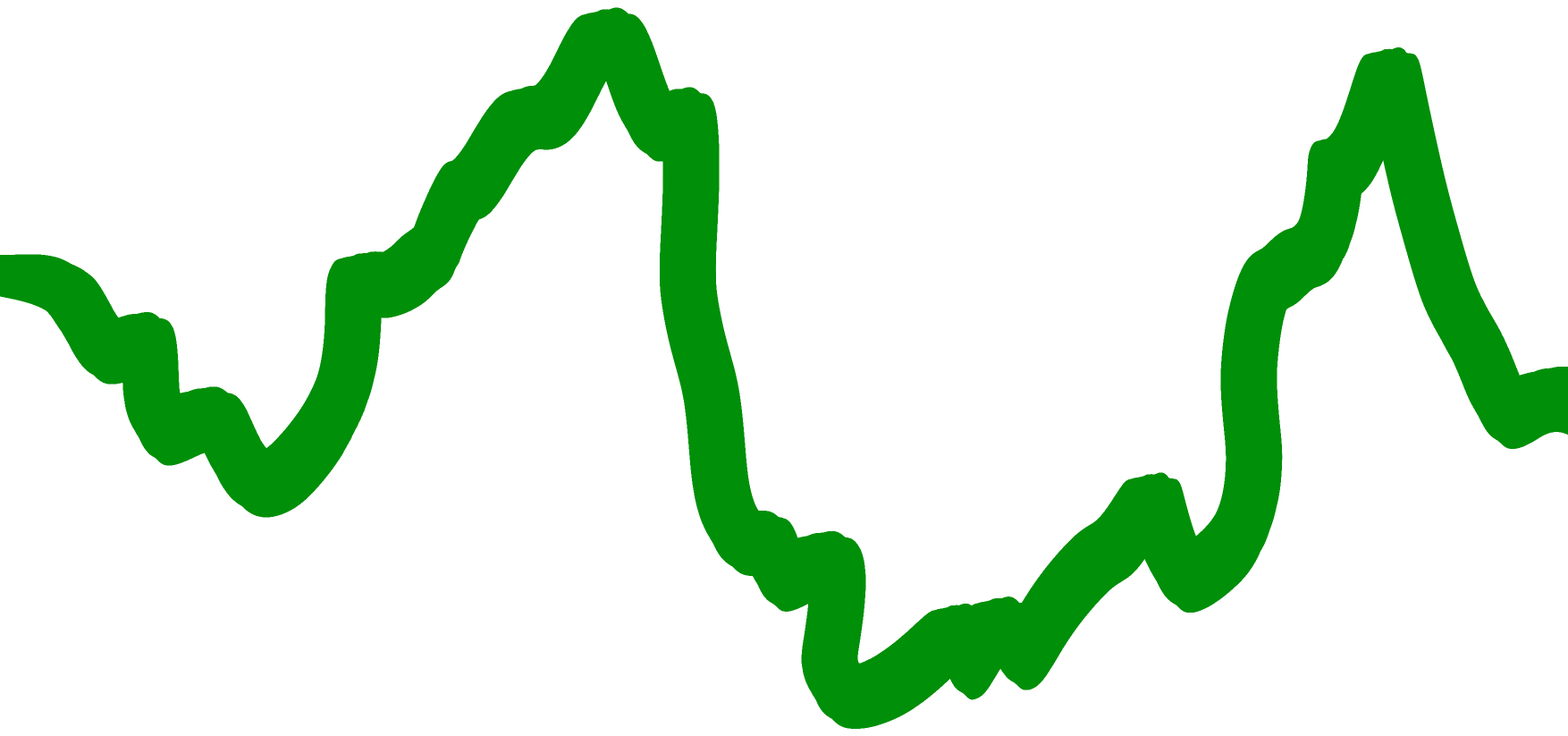}};
 \draw[orange, line width = 0.5 mm] (4.9,3.5) -- (3.75, 3.5);
 \draw[orange, line width = 0.5 mm] (4.9,0.25) -- (4.7, 0.25);
 \draw[orange, line width = 0.5 mm] (4.9,0.25) -- (4.9, 3.5);
 \draw[orange, line width = 0.5 mm] (0.5,3.5) -- (2.75,3.5);
 \draw[orange, line width = 0.5 mm] (0.5,3.5) -- (0.5,2.9);
 \draw[blue, line width = 0.5 mm] (0,0) arc (-90:90:1.5);
 \draw[blue, line width = 0.5 mm] (-1,3) arc (90:270:1.5);
 \draw[black, line width = 0.3 mm] (-1.2,-0.5) -- (2,-0.5);
 \draw[black, line width = 0.3 mm] (2,-0.5) -- (2,1.6);
 \draw[black, line width = 0.3 mm] (2,1.6) -- (2.2,1.6);
 \draw[black, line width = 0.3 mm] (2,1.9) -- (2.2,1.9);
 \draw[black, line width = 0.3 mm] (2.,1.9) -- (2.,3.35);
 \draw[black, line width = 0.3 mm] (2.,3.35) -- (2.85,3.35);
  \draw[ line width =1.2pt] (2.2,1.5) rectangle (3.5,2.1);
  \draw[black, line width = 0.3 mm] (2.75,1.8) arc (0:360:.2);
  \draw[black, line width = 0.3 mm] (3.345,1.8) arc (0:360:.2);
  \draw [gray, line width =0.5pt] (3.13,1.8) -- (3.2,1.95);
 \draw [gray, line width =0.5pt] (2.55,1.8) -- (2.6,1.65);
 \draw[black, line width = 0.3 mm] (-1.2,-0.5) -- (-1.2,0.);
 
 \draw[black, line width = 0.5 mm] (3.9,3.7) arc (45:135:.9);
 \draw[gray, line width = 0.5 mm] (3.63,3.5) arc (45:135:.5);
 %\draw[gray, line width = 0.5 mm] (3.2,3.65) arc (95:135:.5);
  \draw[->, black, line width =0.5 mm]        (3.3,3.3)   -- (3.5,3.9);
 \draw[black, line width = 0.5 mm] (2.95,3.2) -- (3.6, 3.2);
 \draw[black, line width = 0.5 mm] (2.95,3.2) -- (2.65, 3.7);
 \draw[black, line width = 0.5 mm] (3.6,3.2) -- (3.87, 3.7);
 \draw[rounded corners=10pt, line width =1.2pt] (3.2,-.5) rectangle (4.7,1.);
 \draw[rounded corners=10pt, line width =.8pt, color = gray] (3.4,-.3) rectangle (4.5,0.8);
 \node[] at (2.35, 0) {\large $\chi_1$};
 \node[] at (3., 0.2) {\large $r$};
 \node[] at (2.35, 2.7) {\large $\chi_2$};
 %\node[] at (-0.5, .2) {\large System};
 %\node[] at (-0.5, .6) {\large Quantum};

 \begin{axis}[every axis plot post/.append style={color = red,
  line width = 1.8pt, mark=none,domain=-2:2,samples=50,smooth}, % 
  axis line style={opacity=0},
  %All plots: from -2:2, 50 samples, smooth, no marks
  xtick = \empty,
  ytick = \empty,
  yticklabel=\empty,
  xticklabel=\empty,
  axis x line*=none, % no box around the plot, only x and y axis
  axis y line*=left, % the * suppresses the arrow tips
  enlargelimits=upper, at={(-0.24\linewidth,.1\linewidth)}, width=5cm,height=3cm] % extend the axes a bit to the right and top
  \addplot {gauss(0,0.5)};
  %\addplot {gauss(1,0.75)};
\end{axis}
\end{tikzpicture}

%\vspace*{-.3cm}
    \caption{We sketch the schematic of a quantum system coupled to a detector that monitors the system. The readout $r$ obtained due to continuous measurement is noisy. In general, the system is controlled through a parameter $\chi_1$ that changes the system Hamiltonian (unitary control) and another  parameter $\chi_2$ that modifies the measurements (dissipative control) performed on the system.}
    \label{fig-device}
\end{figure}

In this section, we look at the conditional stochastic evolution of a continuously monitored quantum system. Consider a system with Hamiltonian $\hat{H}$\footnote{We assume $\hbar=1$ throughout the analysis.}. Assume the system is coupled to a detector which allows us to monitor an (Hermitian) observable $\hat{L}$ (see Figure~\ref{fig-device}). We further assume that the measurement Kraus operators \cite{BookKraus, BookJordan} are of Gaussian form
\begin{equation}
    \hat{M}(r)=\Big(\frac{d t}{2 \pi \tau}\Big)^\frac{1}{4}\exp\left[-\frac{d t}{4\tau}(r\mathds{1}-\hat{L})^2\right],
    \label{kraus_L}
\end{equation}
where $\tau$ signifies the collapse timescale of the measurement.
The state of the system is expressed by the density matrix $\hat{\rho}$. The Stratonovich stochastic master equation expressing the conditional dynamics of the systems under continuous monitoring is \cite{jacobs2006straightforward, brun_simple_2002}
\begin{equation}
\begin{split}
     \frac{\partial\hat{\rho}}{\partial t}&=\hat{\mathfrak{F}}_{\hat{\rho}}(\hat{\rho}, r)\\&
    =-i[\hat{H},\hat{\rho}] -\frac{1}{4\tau}\left[\Delta \hat{V},\hat{\rho}\right]_++\frac{1}{2\tau}r\left[ \Delta\hat{L}, \hat{\rho}\right]_+,
\end{split}
    \label{SME}
\end{equation}
where we define $\hat{V}=\hat{L}^2$, and $[\cdots,\cdots]_+$ denotes the anticommutator. $\Delta\hat{O}=\hat{O}-\expval{\hat{O}}$, for any operator $\hat{O}$. The subscript $\hat{\rho}$ on $\mathfrak{F}$ denotes the nonlinear nature of the equation. In the subsequent analysis, we will assume that there is a single measurement channel. The results presented can be easily generalized to non-Hermitian observables or multiple measurement channels.
   The readout $r$ obtained in time $dt$ is  
\begin{equation}
    r = \expval{\hat{L}}+\sqrt{\tau}\frac{dW}{dt}.
    \label{r_eq}
\end{equation}
Here, $dW$ is a Wiener noise such that its variance is $dt$ \cite{StochasticCalculus, QuantumNoise}. 

Eq.~\eqref{SME} quantifies the stochastic evolution of the system conditioned on the stochastic readout $r$. Note that Eq.~\eqref{SME} can describe stochastic unravelings of an unmonitored system too
\cite{PhysRevA.46.4363}. In the subsequent sections, we characterize the statistics of the readouts.

\subsection{\label{stoch_H_CDJ}CDJ formalism: Stochastic action and Hamiltonian} 
Now, we provide a brief introduction to the Chantasri-Dressel-Jordan (CDJ) formalism. The CDJ stochastic path integral approach expresses the probability density of realizing a particular quantum trajectory in terms of a path integral of the exponential of a stochastic action \cite{Areeya_Thesis, PhysRevA.88.042110, PhysRevA.92.032125}. This formalism characterizes the statistics of the continuous measurements. By extremizing the stochastic action, the most likely dynamics of the system are obtained. Such most likely dynamics (also called the most likely paths or optimal paths) can be expressed in terms of a Hamiltonian that incorporates both the unitary dynamics and the dissipative dynamics due to measurements.  Thus, we obtain a classical-like description of monitored quantum systems.

In the CDJ stochastic path integral approach, we parametrize Eq.~\eqref{SME}
in terms of coordinates $\pmb{q}$. Here, the coordinates represent the state of the system. For example, they can be the Bloch coordinates for a qubit \cite{PhysRevX.6.041052} or the quadrature expectation values for an oscillator in a coherent state \cite{PRXQuantum.3.010327}.  Assume the initial state of the system is $\pmb{q}(t=0)=\pmb{q}_i$ and the trajectories are postselected such that the final state is $\pmb{q}(t=t_f)=\pmb{q_f}$. The system is monitored  in   $d t$ intervals.  The CDJ stochastic path integral approach expresses the probability density of realizing intermediate states $\{\pmb{q}\}$  and readouts $\{r\}$ in the limit $dt\to 0$ as  \cite{PRXQuantum.3.010327}
\begin{widetext}
\begin{equation}
    \mathcal{P}=\int\mathcal{D}\pmb{p}\:e^\mathcal{S}=\int\mathcal{D} \pmb{p}\: \exp\left[\int_{0}^{t_f}dt(-\pmb{p}\cdot\pmb{\dot{q}}+\mathcal{H}(\pmb{p},\pmb{q},r))\right],
    \label{action_integral}
\end{equation}
\end{widetext}
where $\pmb{p}$ are variables conjugate to $\pmb{q}$. $\mathcal{H}(\pmb{p},\pmb{q},r)$, also called the stochastic Hamiltonian, is a functional of $\{\pmb{p}, \pmb{q}, r\}$. The stochastic Hamiltonian is of the form
\begin{equation}
    \mathcal{H}(\pmb{q},\pmb{p},r)
    = \pmb{p}^\top\cdot\pmb{\mathcal{F}}(\pmb{q},r)+\mathcal{G}(\pmb{q},r),
\label{stoch_h_parametrized}
\end{equation}
where $\dot{\pmb{q}}=\mathcal{F}(\pmb{q},r)$ is the stochastic master equation \eqref{SME}, expressed in terms of the coordinates $\pmb{q}$. The second term, $\mathcal{G}(\pmb{q},r)$ relates to the conditional probability of readout $r$ as measurements are performed on the system in state $\pmb{q}$ in interval $dt$. In other words, $P(r|\pmb{q})\sim e^{d t \mathcal{G}(\pmb{q},r)}$. The most likely paths  (MLP) or optimal paths (OP) can be found by variational extremization of the stochastic action.  The equations for the optimal paths are given by \cite{PhysRevA.88.042110} 
\begin{equation}
\partial_{r}\mathcal{H}=0,\quad \pmb{\dot{q}}=\partial_{\pmb{p}}\mathcal{H},\quad
    \pmb{\dot{p}}=-\partial_{\pmb{q}}\mathcal{H}.
    \label{OP_eq_parametrized}
\end{equation}
The above equation provides a classical-like description of the optimal paths under continuous measurements. Thus, like position and linear momentum in the variables $\pmb{q}$ and $\pmb{p}$ are conjugates of each other. Furthermore, the variables $\pmb{p}$ can be interpreted to quantify the effective `force' a monitored quantum system experiences. See Ref.~\cite{PRXQuantum.3.010327} for a similar description for monitored harmonic oscillators. Define the optimal readout obtained from the first equation above to be $r^\star(\pmb{p}, \pmb{q})$. The second equation then satisfies $\dot{\pmb{q}}=\mathcal{F}(\pmb{q},r^\star)$, consistent with the stochastic master equation.  To extremize the stochastic action with respect to readouts, the stochastic readouts are replaced with a piecewise linear approximation.  Due to such substitution, the optimal readouts obtained are smooth, and the corresponding optimal paths are solutions of an ordinary differential equation. To ensure such approximation is consistent with trajectories obtained from the stochastic master equation, the ordinary differential equations describing the optimal paths must be of the same form as the Stratonovich version of the stochastic master equation (see Wong-Zakai theorem in Refs.~ \cite{wong_convergence_1965, eugene_relation_1965}). The CDJ formalism has been applied to monitored qubits \cite{weber_mapping_2014, jordan_anatomy_2016}, entangled qubits \cite{PhysRevX.6.041052}, monitored oscillators in general Gaussian states \cite{PRXQuantum.3.010327}.

\subsection{\label{stoch_H_General}Stochastic action and Hamiltonian for general continuously monitored systems}The construction of the CDJ stochastic action requires explicit parametrization of the system in terms of a finite number of variables. This can be difficult for many-body systems with very high dimensions or continuous variable systems in non-Gaussian states. In the following formulation, we bypass the need for such parametrization by expressing the most likely evolution in terms of the density matrix.

Eqs.~\eqref{OP_eq_parametrized} can be interpreted as the extremization of the readout probabilities constrained on the conditional stochastic evolution of $\pmb{q}$ \cite{lewalle2023optimal}. The auxiliary variables $\pmb{p}$ in Eq.~\eqref{stoch_h_parametrized} act as Lagrange multipliers.  We utilize these ideas to construct the stochastic Hamiltonian and action for a general continuously monitored system. If the system state at the beginning of a measurement is $\hat{\rho}$, the probability density of obtaining readout $r$ can be expressed as
\begin{equation}
P(r|\hat{\rho})=\textrm{Tr}\left(\hat{M}(r)\hat{\rho}\hat{M}^\dagger(r)\right),
\label{prob_general}
\end{equation}
where $\hat{M}(r)$ is defined in Eq.~\eqref{kraus_L}. We consider the weak measurement limit $dt\ll \tau$ such that the probability density in Eq.~\eqref{prob_general} can be approximated as 
\begin{widetext}
 \begin{equation}
P(r|\hat{\rho})\approx\Big(\frac{d t}{2 \pi \tau}\Big)^\frac{1}{2}\left[1-\frac{dt}{2\tau}\left(r^2-2r\expval{\hat{L}}+\expval{\hat{L}^2}\right)\right].
\label{prob_general_approx}
\end{equation}   
\end{widetext}
Like the auxiliary variables $\pmb{p}$ we introduce a Hermitian operator $\hat{\sigma}$ and  
generalize Eq.~\eqref{stoch_h_parametrized} to define the stochastic Hamiltonian as \cite{tkthesis}
\begin{equation}
\mathcal{H}dt\sim\textrm{Tr}\left(\hat{\sigma}\hat{\mathfrak F}_{\hat{\rho}}(\hat{\rho},r)dt\right)+\ln P(r|\hat{\rho}).
\label{H_stoch_eq}
\end{equation}
The operator $\hat{\sigma}$ is called the costate. Note, the stochastic Hamiltonian in Eq.~\eqref{H_stoch_eq} is currently an ansatz. We will shortly see that this formulation produces the CDJ most likely paths. In the Appendix \ref{CDJ_vs_costate}, we show (for a single qubit system) that such a costate-based description leads to a stochastic Hamiltonian identical to Eq.~\eqref{stoch_h_parametrized}.   Furthermore, in Section \ref{text_pontrya} we will see how the stochastic Hamiltonian in Eq.~\eqref{H_stoch_eq} is a special case of Pontryagin's maximum principle. As mentioned previously, costate-based description for the optimal control of open quantum systems has been adopted in the past \cite{jordan_anatomy_2016,Koch_2016, PhysRevApplied.16.054023}.
Ignoring multiplicative constants and using the approximation in Eq.~\eqref{prob_general_approx},  the stochastic Hamiltonian takes the following form
\begin{widetext}
\begin{equation}
\mathcal{H}(\hat{\sigma},\hat{\rho}, r) =\textrm{Tr}\left(\hat{\sigma}\hat{\mathfrak F}_{\hat{\rho}}(\hat{\rho},r)\right)-\frac{1}{2\tau}\left(r^2-2r\expval{\hat{L}}+\expval{\hat{L}^2}\right).
\label{stoch_h_general}
\end{equation}
\end{widetext}
The stochastic action can be written as
\begin{equation}
\mathcal{S} =\int_0^{t_f}dt\left(-\textrm{Tr}\left(\hat{\sigma}\frac{\partial\hat{\rho}}{\partial t}\right)+\mathcal{H}\right).
\label{stoch_S_general}
\end{equation}
The most likely path under continuous measurements, also called the optimal path, can be found by the variational extremization of the stochastic action $\delta \mathcal{S}=0$.  

Then, the optimal readout is given by  (see Appendix \ref{appA0_strat})
\begin{equation}
    r^\star =\expval{\hat{L}}+ \frac{1}{2}\expval{\Big[\Delta\hat{L},\hat{\sigma}\Big]_+}=\expval{\hat{L}}+u.
    \label{opt_readout}
\end{equation}
The first term denotes the signal. The second term $u = \frac{1}{2}\expval{\Big[\Delta\hat{L},\hat{\sigma}\Big]_+}$ denotes the optimal noise.  The equations for the optimal path are given by (see Appendix \ref{appA0_strat})
\begin{equation}
    \begin{split}
        \frac{\partial\hat{\rho}}{\partial t}&=-i[\hat{H},\hat{\rho}] -\frac{1}{4\tau}\Big[\Delta\hat{V},\hat{\rho}\Big]_++\frac{r^\star}{2\tau}\Big[\Delta\hat{L},\hat{\rho}\Big]_+,\\\frac{\partial\hat{\sigma}}{\partial t}&= -i[\hat{H},\hat{\sigma}] +\frac{1}{4\tau}\left[\Delta\hat{V},\hat{\sigma}\right]_+-\frac{r^\star}{2\tau}\Big[\Delta\hat{L},\hat{\sigma}\Big]_+.
    \end{split}
    \label{op_eqns}
\end{equation}
where we choose $\expval{\hat{\sigma}}=1$. We can always make this choice since Eq.~\eqref{SME} preserves the trace. Therefore, adding a scalar to $\hat{\sigma}$ does not change the stochastic Hamiltonian in Eq.~\eqref{stoch_h_general}. Note that the costate evolution along the optimal path is given by the negative conjugate of the state evolution. In other words,
\begin{equation}
    \frac{\partial\hat{\sigma}}{\partial t} = -\hat{\mathfrak F}^{\dagger}_{\hat{\rho}}(\hat{\sigma},r^\star),
\end{equation}which may be interpreted as the time reversal of the state evolution. This is consistent with past work using the costate formalism \cite{Koch_2016}.   We will shortly derive a more general expression for the costate evolution where such correspondence might not be valid.
The optimal Hamiltonian $\mathcal{H}^\star(\hat{\sigma},\hat{\rho})=\mathcal{H}(\hat{\sigma},\hat{\rho},r^\star)$ is given by
\begin{equation}
    \begin{split}
\mathcal{H}^\star(\hat{\sigma},\hat{\rho}) &=\frac{1}{8\tau}\expval{\left[\Delta\hat{L},\hat{\sigma}\right]_+}^2\\&+\expval{i[\hat{H},\hat{\sigma}]}-\frac{1}{4\tau}\expval{\left[(\Delta\hat{L})^2,\hat{\sigma}\right]_+}\\&= \expval{i[\hat{H},\hat{\sigma}]}\\&+\frac{1}{8\tau}\expval{\left[\hat{L},\hat{\sigma}\right]_+}^2-\frac{1}{4\tau}\expval{\left[\hat{V},\hat{\sigma}\right]_+}.
    \end{split}
    \label{h_op_general}
\end{equation} Equations \eqref{op_eqns} are the equations for the optimal path of a general continuously monitored system. Such formalism can be applied to finite-dimensional systems such as qubits \cite{RevModPhys.93.025005}, qudits \cite{10.3389/fphy.2020.589504}, many-body systems \cite{fazio2025manybodyopenquantumsystems}, and also to continuous variable systems \cite{RevModPhys.77.513}, multimode physics in ultrafast optics \cite{Yanagimoto:22, Yanagimoto:24}.  

Thus, we have provided a costate-based description of the CDJ most likely paths. Our subsequent analysis focuses on the monitored harmonic oscillator.
\subsection{\label{sho}Example: optimal paths of a continuously monitored quantum harmonic oscillator}
We consider a quantum harmonic oscillator with  Hamiltonian\footnote{In the subsequent analysis, the time and the collapse timescale $\tau$ are in units of $1/\omega$, i.e., in terms of the inverse of the oscillator frequency. The quadratures are made dimensionless by scaling position and momentum by $\sqrt{m\omega/
\hbar}$ and $\frac{1}{\sqrt{\hbar m \omega}}$ respectively (see Ref.~\cite{PRXQuantum.3.010327}).  Then, the stochastic readout is dimensionless as well.}.
\begin{equation}
    \hat{H}=\frac{1}{2}\left(\hat{X}^2+\hat{P}^2\right),
\label{hamiltonian_sho}
\end{equation}
where $\hat{X}$ and $\hat{P}$ are  dimensionless position and momentum observables. We consider adaptive measurements of the quadrature given by 
\begin{equation}
    \hat{L}_\theta=\cos\theta(t)\hat{X}+\sin\theta (t)\hat{P},
    \label{l_theta}
\end{equation}
where $\theta(t) (\equiv \chi_2(t))$ is in general a dissipative control parameter. The conjugate quadrature is
\begin{equation}
    \hat{M}_\theta=-\sin\theta(t)\hat{X}+\cos\theta (t)\hat{P}.
    \label{m_theta}
\end{equation}
Such measurements are realized by clamping a mechanical oscillator in a cavity and performing a balanced homodyne detection on the transmitted light \cite{RevModPhys.86.1391,RevModPhys.82.1155, rossi_measurement-based_2018, PhysRevLett.123.163601}. The measurement collapse timescale $\tau$ in this case is the inverse of measurement rate and depends on the cavity decay rate, cavity resonator optomechanical field enhanced coupling strength, and detection efficiency \cite{tkthesis,PhysRevLett.123.163601}. In this article, we assume completely efficient detectors.

The Stratonovich stochastic master equation for the conditional   evolution of the monitored oscillator is given by 
\begin{equation}
\begin{split}
     &\frac{\partial\hat{\rho}}{\partial t}
    =\hat{\mathfrak{F}}_\theta(\hat{\rho},r)\\&=-i[\hat{H},\hat{\rho}] -\frac{1}{4\tau}\Big[\Delta\hat{V}_\theta,\hat{\rho}\Big]_++\frac{1}{2\tau}r\Big[\Delta\hat{L}_\theta,\hat{\rho}\Big]_+,
\end{split}
    \label{SME_SHO}
\end{equation}
with the readout $r=\expval{\hat{L}_\theta}+\sqrt{\tau}\frac{dW}{dt}$. The optimal paths for the monitored harmonic oscillator in general Gaussian states have been analyzed previously \cite{PRXQuantum.3.010327}. Subsequent analysis 
looks at the optimal path for general states.
\subsection{\label{sho_op_strato}Optimal readout and paths}

Following the formulation presented in Sec.~\ref{stoch_H_CDJ}, we now characterize the most likely readouts and path of the monitored oscillator. The expression for the optimal readout is 
\begin{equation}
    r^\star=r_\theta=\frac{1}{2}\expval{\Big[\hat{L}_\theta,\hat{\sigma}\Big]_+}.
    \label{op_readout_sho}
\end{equation}
The equations for the optimal paths are
\begin{subequations}
\begin{align}
    \begin{split}
        \frac{\partial\hat{\rho}}{\partial t}&=-i[\hat{H},\hat{\rho}] -\frac{1}{4\tau}\Big[\Delta\hat{V}_\theta,\hat{\rho}\Big]_++\frac{1}{2\tau}r_\theta\Big[\Delta\hat{L}_\theta,\hat{\rho}\Big]_+,
        \label{op_rho_sho}
        \end{split}\\
        \begin{split}
        \frac{\partial\hat{\sigma}}{\partial t}&=-i[\hat{H},\hat{\sigma}] +\frac{1}{4\tau}\Big[\Delta\hat{V}_\theta,\hat{\sigma}\Big]_+-\frac{1}{2\tau}r_\theta\Big[\Delta\hat{L}_\theta,\hat{\sigma}\Big]_+.\label{op_sigma_sho}
    \end{split}
    \end{align}
    \label{op_sho}
\end{subequations}We see that  $\hat{\sigma}$ influences the evolution of the state only through the optimal readout $r_\theta$, which is a real scalar. We can reduce the problem's dimensionality significantly by looking at the evolution of $r_\theta$ only. Similar to Eq.~\eqref{op_readout_sho}, we define the scalar $v_\theta$ in terms of the conjugate quadrature as
\begin{equation}
   v_\theta =  \frac{1}{2}\expval{\left[\hat{M}_\theta,\hat{\sigma}\right]_+}.  
\end{equation}We also adopt the following definitions,
\begin{equation}
        \quad w_\theta = i\expval{\left[\hat{\sigma},\hat{L}_\theta\right]},\quad z_\theta = i\expval{\left[\hat{\sigma},\hat{M}_\theta\right]}.
\end{equation}
We can now express the evolution of the optimal readout in terms of the following coupled equations (see Appendix \ref{app:cdjp_SHO})
\begin{subequations}
    \begin{align}
        \begin{split}
    &\frac{dr_\theta}{dt}=\dot{\phi}v_\theta, \quad \frac{dv_\theta}{dt}=-\dot{\phi}r_\theta+\frac{w_\theta}{4\tau},
    \label{u_eq}
        \end{split}\\
\begin{split}
   \frac{dw_\theta}{dt}=\dot{\phi}z_\theta,\quad \frac{dz_\theta}{dt}=-\dot{\phi}w_\theta,
   \label{wz_eq}
\end{split}
    \end{align}
\label{uvwz_eq}
\end{subequations}where $\dot{\phi}=\dot{\theta}+1$. Note, the $\dot{\theta}$ in the definition of $\dot{\phi}$ arises due to the varying quadrature measurement. The additional 1 from the oscillator's unitary dynamics. By integrating Eqs.~\eqref{op_rho_sho} and \eqref{uvwz_eq} together, we can find the optimal paths of a monitored harmonic oscillator in arbitrary states. This simplification is advantageous in numerical simulations since, instead of two operators $\hat{\rho}$ and $\hat{\sigma}$, we only need to integrate one operator $\hat{\rho}$ and four scalars.
This analysis goes beyond the Gaussian state assumption in Ref.~\cite{PRXQuantum.3.010327}.

\begin{figure}[!t]
	\includegraphics[width=0.90\linewidth,trim = {0 0 0 0}, clip]{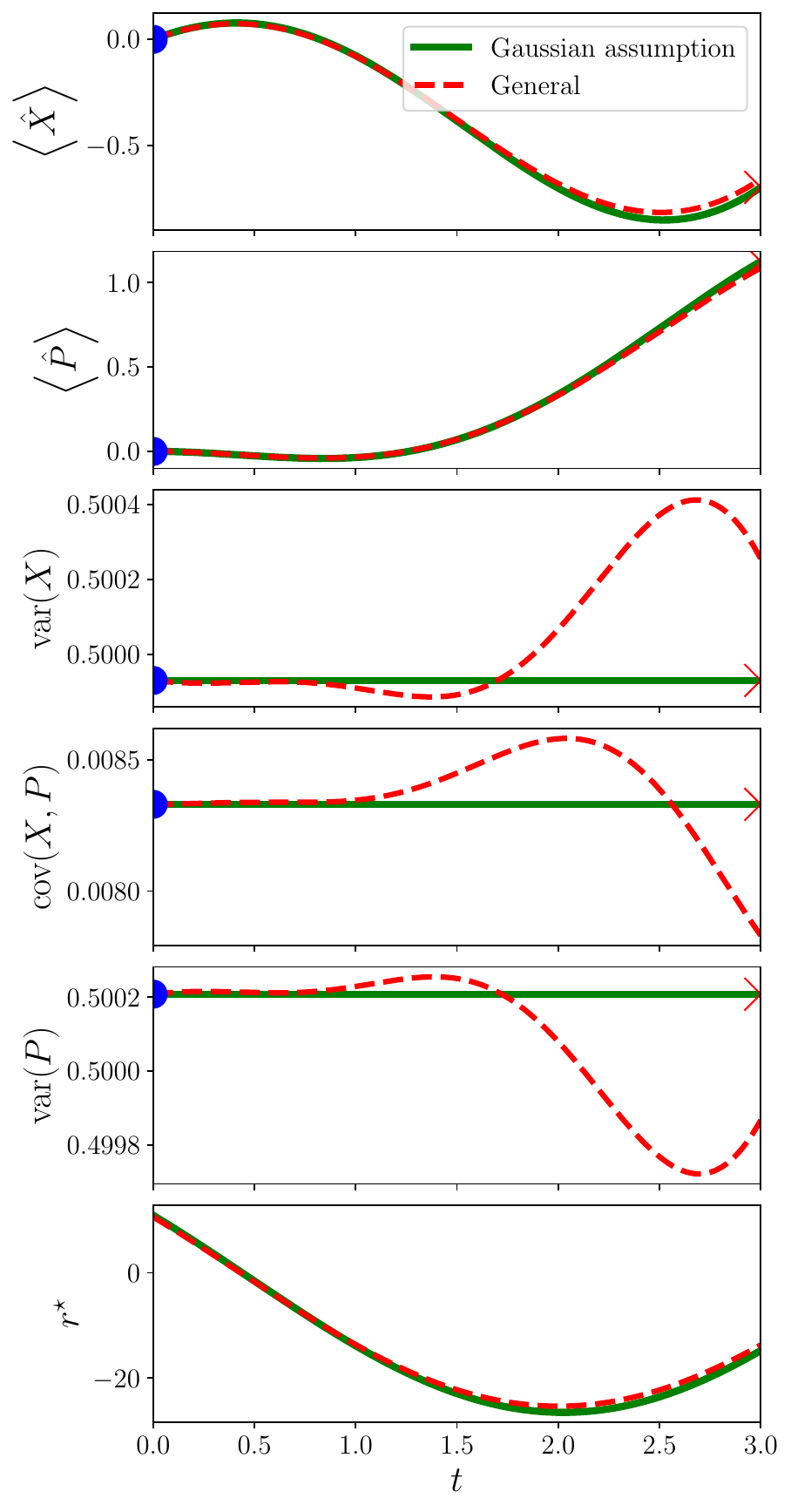}
	\caption{The plots show the optimal path for Gaussian states under position measurements (i.e.~$\theta=0$ in Eq.~\ref{l_theta}). The top two panels show the time evolution of the expectation values of position and momentum. The next three panels show the time evolution of the position variance, position-momentum covariance, and the momentum variance.  The final panel shows the most likely readout obtained as a function of time. All the observables are scaled to be dimensionless (see Sec.~\ref{sho}). The time and the collapse timescale ($\tau=15.0$) are in units of the inverse of the oscillator frequency.   In the plots, the green curve shows evolution obtained from Ref.~\cite{PRXQuantum.3.010327}, with a Gaussian state assumption. The red dashed lines show the most likely path obtained from the general formalism presented in  Sec.~\ref{sho_op_strato}. The blue dots in the first five panes denote the initial (a squeezed vacuum state), and the red crosses denote the final state (a squeezed coherent state).  The squeezing is determined by the $\tau$ (see Appendix \ref{xmeasurement}). We see that the expectation values match very well, while the deviations in the variances and the covariance signify the failure of the steady-state assumption.}
	\label{tmpfig}%
\end{figure}

We assume $\phi(t)=\theta(t)+t$ and $\phi(0)=\theta(0)=0$ without the loss of generality.  Eqs.~\eqref{wz_eq} can be integrated to find
\begin{equation}
\begin{split}
    w_\theta(t)&=A\cos\phi(t)+B\sin\phi(t),\\
    z_\theta(t)& = -A\sin\phi(t)+B\cos\phi(t),
\end{split}
\end{equation}
where $A$ and $B$ are real constants corresponding to the initial values of $w_\theta$ and $z_\theta$. Putting these values into Eqs.~\eqref{u_eq}, we can solve for $u_\theta$ and $v_\theta$ as
\begin{equation}
\begin{split}
    r_\theta(t)+iv_\theta(t)&=e^{-i\phi(t)}\times\\&\left(\alpha+\frac{it}{8\tau}(A+iB)+\int_0^tdt^\prime h(t^\prime)\right),
\end{split}
\label{r_th_expr}
\end{equation}
with 
\begin{equation}
    h(t)=\frac{i}{8\tau}(A-iB)e^{i2\phi(t)},
\end{equation}
and $\alpha$ is a complex number. Solving for the optimal path then involves integrating Eqs.~\eqref{op_rho_sho} and \eqref{r_th_expr}  from the initial state $\hat{\rho}(0)=\hat{\rho}_i$ and finding the values of $\alpha$, $A$, and $B$ such that $\hat{\rho}(t_f)=\hat{\rho}_f$. Thus, we have solved for the optimal paths of a monitored simple harmonic oscillator in an arbitrary state. According to Eqs.~\eqref{uvwz_eq}, the optimal readout behaves like a resonantly driven oscillator with frequency $\dot{\phi}$. This is consistent with optimal path solutions previously found for Gaussian states \cite{PRXQuantum.3.010327}.    Figure~\ref{tmpfig} shows that our formalism for general optimal paths reproduces previous results of  Gaussian state evolution in Ref.~\cite{PRXQuantum.3.010327} (see Appendix \ref{numerics_theta0} for details on the numerical method). Here, the control parameter $\theta(t)=0$. Note that the covariance matrix elements in the general framework show a slight deviation from the final state.  Due to a steady state assumption (see Ref.~\cite{PRXQuantum.3.010327}), the variances and covariances do not evolve in time for the Gaussian state formalism. The numerical solution to the general optimal path produces a final state that is very close to the target state. However, the variances and the covariance for the general optimal path deviate slightly from the final state since position measurements alone are insufficient to attain the target state. Such deviations signify the failure of the steady-state assumption made in Ref.~\cite{PRXQuantum.3.010327}. The question of the reachability of specific states under available control is interesting, although it falls beyond the scope of our current endeavor.

Thus, the costate-based analysis of the CDJ most likely paths reproduces the previous coordinate-based results for the monitored oscillator in general Gaussian states. In the subsequent analysis, we will show that the CDJ most likely paths can be derived as a special case of a general Pontryagin's maximum principle.

\section{\label{text_pontrya}Pontryagin maximum principle for the optimal control of  quantum systems}
We now provide a formal statement of the Pontryagin maximum principle for arbitrary quantum systems. We show that the most likely paths provided by the Chantasri-Dressel-Jordan stochastic path integral formalism can be derived as a special case of the following analysis.  We consider quantum systems in state $\hat{\rho}(t)$, with an evolution characterized by 
\begin{equation}
    \frac{\partial \hat{\rho}}{\partial t}=\mathcal{F}_{\hat{\rho},\chi}[\hat{\rho}],
    \label{gen_rho_evol}
\end{equation}
where the superoperator $\mathcal{F}_{\hat{\rho},\chi}$ depends on control $\chi(t)$ and the state $\hat{\rho}(t)$. Thus, Eq.~\eqref{gen_rho_evol} is, in general, nonlinear. Further, assume $\mathcal{F}_{\hat{\rho},\chi}$ depends on $\hat{\rho}$ through expectation values of different observables such that, 
\begin{widetext}
    \begin{equation}
    \mathcal{F}_{\hat{\rho}+\delta\hat{\rho},\chi}[\hat{\rho}+\delta\hat{\rho}]=\mathcal{F}_{\hat{\rho},\chi}[\hat{\rho}]+\mathcal{F}_{\hat{\rho},\chi}[\delta\hat{\rho}]+\sum_{\nu=0}^{\nu_1}\textrm{Tr}(\hat{A}_{\nu,\chi}\delta\hat{\rho})\mathcal{F}^\nu_{\hat{\rho},\chi}[\hat{\rho}],
    \label{eq_F_variation}
\end{equation}
\end{widetext}

up to first order in $\delta \hat{\rho}$. Here $\hat{A}_{\nu,\chi}$ are Hermitian observables and $\mathcal{F}^\nu_{\hat{\rho},\chi}$ are superoperators for $\nu= 0,\cdots, \nu_1$.

For continuously monitored systems, we consider a continuous approximation of the readout $r$, and Eq.~\eqref{gen_rho_evol} corresponds to the Stratonovich master equation \cite{eugene_relation_1965, wong_convergence_1965}. Without loss of generality,  we mathematically treat the readout as a part of the control in this section.

We consider system evolving from $\hat{\rho}(t=t_0)=\hat{\rho}_0$ to $\hat{\rho}(t=t_1)=\hat{\rho}_f$ and introduce the following cost functional that will be minimized
\begin{equation}
    J=\int_{t_0}^{t_1}f^0(\hat{\rho}(t),\chi(t))dt,
    \label{gen_cost_fn}
\end{equation}
where  $t_1$ is not fixed. $f^0$ is an as-yet-unspecified function of the system state and the control.
We can express the variation of $f^0$ as follows
\begin{equation}
    f^0(\hat{\rho}+\delta \hat{\rho},\chi)=f^0(\hat{\rho},\chi)+\textrm{Tr}\left(\frac{\delta f^0}{\delta \hat{\rho}}\delta\hat{\rho}\right).
\end{equation}
We assume the controls are $\chi:[t_0,t_1]\to U_{\textrm{C}}$ with $U_{\textrm{C}}\subset \mathbb{R}^m$ such that the set of $\chi(t)$ for $t_0\leq t\leq t_1$ has a compact closure in $\mathbb{R}^m$ (i.e.~bounded) \cite{pontryaginbook}. In this work, the control is assumed to be piecewise continuous. However, the results apply to a more general class of controls (measurable and bounded, see Ref.~\cite{pontryaginbook}). We also assume the cost is a smooth function of the quantum state and the control variables.

\emph{Problem statement}: For all possible controls $\chi(t)$ that take the system from $\hat{\rho}_0$ to $\hat{\rho}_f$, find the one $\chi^\star(t)$ that minimizes the cost in Eq.~\eqref{gen_cost_fn}.

We can look at the problem geometrically. We define the variable 
\begin{equation}
    q^0(t)=\int_{t_0}^{t}f^0(\hat{\rho}(t),\chi(t))dt,
\end{equation}
and an extended coordinate $Q(t)\equiv\left(q^0(t),\hat{\rho}(t)\right)$. Then, the cost in Eq.~\eqref{gen_cost_fn} is given by $J=q^0(t_1)$. Therefore,  we consider all controls $\chi(t)$ that starts from the coordinate $Q(t_0)=(0,\hat{\rho}_0)$ and and ends up at the line  $Q(t_1)=(q^0(t_1),\hat{\rho}_f)$. Optimal control corresponds to the minimization of the coordinate $q^0(t_1)$ over such controls. We also define the auxiliary variables $P=(p_0,\hat{\sigma})$, where
$p_0$ is a real number and $\hat{\sigma}$ is a Hermitian operator (costate). The auxiliary variables evolve according to 
\begin{equation}
    \frac{dp_0}{dt}=0,
    \label{dp0}
\end{equation}
and 
\begin{equation}
    \frac{\partial \hat{\sigma}}{\partial t}=-p_0\frac{\delta f^0}{\delta \hat{\rho}}-
    \mathcal{F}_{\hat{\rho},\chi}^\dagger[\hat{\sigma}]-\sum_{\nu=0}^{\nu_1}\hat{A}_{\nu,\chi}\textrm{Tr}(\hat{\sigma}\mathcal{F}^\nu_{\hat{\rho},\chi}[\hat{\rho}]).
    \label{dsigma_gen}
\end{equation}We consider the Hamiltonian given by
\begin{equation}
    \mathcal{H}(P,Q,\chi)=p_0f^0+\textrm{Tr}\left(\hat{\sigma}\mathcal{F}_{\hat{\rho},\chi}[\hat{\rho}]\right).
\end{equation} For given $Q$ and $P$, we define the supremum of the above Hamiltonian with respect to the control as
\begin{equation}
    \mathcal{K}(P,Q)=\sup_{\chi\in U_{\textrm C}}\mathcal{H}(Q,P,\chi).
\end{equation}

\emph{The Pontryagin maximum principle} \cite{pontryaginbook}: The necessary condition for control $\chi^\star(t)$ and trajectory $Q(t)$ to be optimal is $\exists$ a non-zero $P(t)$ corresponding to $\chi^\star(t)$ and $Q(t)$ such that\\
(\emph{Condition I}) the function $\mathcal{H}(P(t), Q(t), \chi)$ attains its maximum at $\chi^\star(t)$ for almost all $t_0\leq t\leq t_1$:
\begin{equation}
    \mathcal{H}(P(t),Q(t),\chi^\star(t))=\mathcal{K}(P(t),Q(t)).
    \label{PMP_cond_hmax}
\end{equation}
  (\emph{Condition II}) At the final time, the following conditions are true:
   \begin{subequations}
       \begin{align}
           \begin{split}
               p_0(t_1)&\leq0,\label{PMP_cond_p0}\end{split}\\
       \begin{split}
               \mathcal{K}(P(t_1),Q(t_1))&=0.
           \label{PMP_cond_k0}\end{split}\end{align}
   \end{subequations}
Furthermore, if $P(t)$ satisfies Eqs.~\eqref{dp0} and \eqref{dsigma_gen}, $p_0$ and $\mathcal{K}$ are constants. Thus, Condition II is satisfied for any $t_0\leq t\leq t_1$.

The above principle can be proven by following the steps laid out in Ref.~\cite{pontryaginbook}, with the equations of motion replaced by Eq.~\eqref{gen_rho_evol}. Note that the conditions are necessary but not sufficient. One should verify the existence of optimal control before applying the maximum principle. Also, a solution derived from the maximum principle might not be optimal when multiple such solutions are possible. Further investigations are required to ensure optimality in such scenarios.   If the control $\chi(t)$ is piecewise continuous, like in our considerations, Condition I is satisfied everywhere. Also, for a fixed time interval, the constraint specified by Eq.~\eqref{PMP_cond_k0} is lifted. 

\emph{For variable end points} : The end points $\hat{\rho}_0$ and $\hat{\rho}_f$ might  not be fixed and can belong to manifolds $\mathcal{M}_0$ and $\mathcal{M}_f$ of dimensions $r_0$ and $r_f$, respectively. Assume $\hat{\rho}(t)$ is the corresponding optimal solution. Also, assume $T_{\hat{\rho}(t_0)}\mathcal{M}_0$ and $T_{\hat{\rho}(t_1)}\mathcal{M}_f$ are tangent spaces of $\mathcal{M}_0$ and $\mathcal{M}_f$ at $\hat{\rho}(t_0)$ and $\hat{\rho}(t_1)$ respectively. The tangent spaces have dimensions $r_0$ and $r_f$, too. Then, for optimal control, the following holds:\\
(\emph{Condition III} or the transversality condition) If $\Delta \hat{\rho}_0\in T_{\hat{\rho}(t_0)}\mathcal{M}_0$   and $\Delta \hat{\rho}_f\in T_{\hat{\rho}(t_1)}\mathcal{M}_0$, then the following holds 
     \begin{subequations}
       \begin{align}
           \begin{split}
               \textrm{Tr}(\hat{\sigma}(t_0)\Delta \hat{\rho}_0)&=0,\label{PMP_cond_trans_0}\end{split}\\
       \begin{split}
               \textrm{Tr}(\hat{\sigma}(t_f)\Delta \hat{\rho}_f)&=0.
           \label{P_cond_trans_f}\end{split}\end{align}
   \end{subequations}
We will only consider the fixed-endpoint problems in this article for simplicity.

Therefore, we described a general Pontryagin's maximum principle for quantum systems undergoing arbitrary dynamics.
\subsection{Optimal control in continuously monitored systems}Now, we apply the above results to continuously monitored systems (see Refs.~\cite{kokaew_quantum_2026, lewalle2023optimal}) with a stochastic master equation given by, 
\begin{equation}
    \frac{\partial \hat{\rho}}{\partial t}=\hat{\mathfrak{F}}_{\hat{\rho}}(\hat{\rho},r),
    \label{rho_evol_append}
\end{equation}
where $r$ is the readout. We will examine most likely path based optimal controls for such monitored systems. We refer to the most likely path based optimal control framework as the CDJP (CDJ-Pontryagin) formulation, consistent with the description in Ref.~\cite{lewalle2023optimal}. We consider trajectories with initial state $\hat{\rho}(0)=\hat{\rho}_0$ and final state $\hat{\rho}(t_f)=\hat{\rho}_f$. We divide the total interval $[0,t_f]$ into small intervals of length $\Delta t$ such that $n\Delta t=t_f$. Assume we obtain readouts $r_0, r_1, \cdots r_{n-1}$ in each $\Delta t$ interval. Using Eq.~\eqref{rho_evol_append}, we can update the corresponding quantum state conditionally to $\hat{\rho}_0, \hat{\rho}_1,\cdots, \hat{\rho}_{n-1}, \hat{\rho}_n=\hat{\rho}_f$.   Then the probability density of obtaining the set of readouts $\{r_k\}=r_0, r_1, \cdots r_{n-1}$, with initial state   $\hat{\rho}_0$ is given by \cite{PhysRevA.88.042110, PRXQuantum.3.010327} 
\begin{equation}
\begin{split}
    P(r_{n-1},
    \cdots,& r_2, r_1,r_0|\hat{\rho}_0)=\\&P(r_{n-1},
    \cdots, r_2, r_1|r_0,\hat{\rho}_0)P(r_{0},
    |\hat{\rho}_0).
\end{split}
\end{equation}
Now, from Eqs.~\eqref{prob_general} and \eqref{prob_general_approx}, we can approximate (up to first order in $\Delta t$) the second term on the right hand side as 
\begin{equation}
    P(r_0|\hat{\rho}_0)\approx \Gamma_{\Delta t} e^{-\frac{\Delta t}{2\tau}\left(r_0^2-2r_0\expval{\hat{L}}+\expval{\hat{L}^2}\right)},
\end{equation}
where $\Gamma_{\Delta t}$ is a constant. The condition of the initial state $\hat{\rho}_0$ and obtaining readout $r_0$ in the first term is the same as considering the state $\hat{\rho}_1$ at $t=\Delta t$ for Markovian dynamics. Thus, we can write 
\begin{widetext}
    \begin{equation}
%\begin{split}
    P(r_{n-1},
    \cdots, r_2, r_1,r_0|\hat{\rho}_0)=P(r_{n-1},
    \cdots, r_2, r_1|\hat{\rho}_1)\times \Gamma_{\Delta t} e^{-\frac{\Delta t}{2\tau}\left(r_0^2-2r_0\expval{\hat{L}}+\expval{\hat{L}^2}\right)}.
%\end{split}
\end{equation}
\end{widetext}

Carrying out such decomposition and taking the limit $\Delta t\to 0$, the probability density of obtaining a specific set of readouts (thus, realizing a specific trajectory) is
\begin{equation}
    P(r(t)|\hat{\rho}_0)=\tilde{\Gamma}_{\Delta t}e^{-\int_0^{t_f}\frac{d t}{2\tau}\left(r^2-2r\expval{\hat{L}}+\expval{\hat{L}^2}\right)},
\end{equation}
which is a functional of the trajectory. The most likely trajectory is obtained by maximizing the argument of the exponent or, equivalently, minimizing the functional 
\begin{equation}
    J=\int_0^{t_f}\frac{d t}{2\tau}\left(r^2-2r\expval{\hat{L}}+\expval{\hat{L}^2}\right).
    \label{Jcost}
\end{equation}
Thus, for continuously monitored systems, we can identify the cost function in Eq.~\eqref{gen_cost_fn} as
\begin{equation}
    f^0(\hat{\rho}, r, \chi)=\frac{1}{2\tau}\left(r^2-2r\expval{\hat{L}_\chi}+\expval{\hat{L}_\chi^2}\right),
\end{equation}
where we take the possible control $\chi$ dependence of the monitored observable into account. To apply Pontryagin's maximum principle, we construct the stochastic Hamiltonian
\begin{equation}
    \mathcal{H}(p_0,\hat{\sigma}, \hat{\rho}, r, \chi)=p_0f^0(\hat{\rho}, r, \chi)+\textrm{Tr}(\hat{\sigma}\hat{\mathfrak{F}}(\hat{\rho},r)).
    \label{eq_H_pontrya_r}
\end{equation}
From Condition II and Eq.~\eqref{dp0}, we see that $p_0\leq 0$ and $p_0= const.$ In this article, we ignore the abnormal optimal ($p_0=0$) therefore, $p_0<0$. Note that multiplying $\mathcal{H}$ by a constant only changes the adjoint coordinate $P(t)=(p_0,\hat{\sigma})$ by a constant factor. There is no change in the most likely path evolution and the optimal control conditions. Hence, we can choose $p_0=-1$ such that the stochastic Hamiltonian becomes 
\begin{equation}
    \mathcal{H}(\hat{\sigma}, \hat{\rho}, r, \chi)=-f^0(\hat{\rho}, r, \chi)+\textrm{Tr}(\hat{\sigma}\hat{\mathfrak{F}}(\hat{\rho},r)).
    \label{H_pontrya_ctsms}
\end{equation}
In this case, comparing the variation of the stochastic master equation with Eq.~\eqref{eq_F_variation}, we get the following equations
\begin{equation}
\begin{split}
    \hat{A}_{0,\chi}&=\frac{1}{2\tau}\hat{L}_\chi^2,\quad \mathcal{F}^0_{\hat{\rho},\chi}=\hat{\mathds{1}},\\ \hat{A}_{1,\chi}&=-\frac{r}{\tau}\hat{L}_\chi,\quad \mathcal{F}^1_{\hat{\rho},\chi}=\hat{\mathds{1}},
\end{split}
\end{equation}
where $\hat{\mathds{1}}$ denotes the identity operator. Also, from the variation of $f^0$ we obtain the following \begin{equation}
    \frac{\partial f^0}{\partial \hat{\rho}}=\frac{1}{2\tau}\left(-2r\hat{L}_\chi+\hat{L}_\chi^2\right).
\end{equation}
Thus, the costate evolution from Eq.~\eqref{dsigma_gen} takes the form given by
\begin{equation}
\begin{split}
    \frac{\partial \hat{\sigma}}{\partial t} =&-i[\hat{H}_\chi,\hat{\sigma}] +\frac{1}{4\tau}\left(\left[\Delta\hat{V}_\chi,\hat{\sigma}\right]_+-2\expval{\hat{\sigma}}\hat{V}_\chi\right)\\&-\frac{r}{2\tau}\left(\Big[\Delta\hat{L}_\chi,\hat{\sigma}\Big]_+-2\expval{\hat{\sigma}}\hat{L}_\chi\right)\\&+\frac{1}{2\tau}(-2r\hat{L}_\chi+\hat{L}_\chi^2),
\end{split}
\end{equation}
which is the same as Eqs.~\eqref{gamma12_opt} and \eqref{OPeq_appnx} in Appendix~\ref{appA0_strat}, where the costate evolution has been derived from the variation of the stochastic action.
Since Eq.~\eqref{rho_evol_append} is trace-preserving, we can redefine $\hat{\sigma}=\hat{\sigma}+(1-\expval{\hat{\sigma}})\hat{\mathds{1}}$ without changing the value of the stochastic  Hamiltonian in Eq.~\eqref{H_pontrya_ctsms}. With this redefinition, the evolution of the costate becomes 
\begin{equation}
\begin{split}
    \frac{\partial \hat{\sigma}}{\partial t} &= -\hat{\mathfrak{F}}_{\hat{\rho}}^\dagger(\hat{\sigma},r)\\=&-i[\hat{H}_\chi,\hat{\sigma}] +\frac{1}{4\tau}\left[\Delta\hat{V}_\chi,\hat{\sigma}\right]_+-\frac{r}{2\tau}\Big[\Delta\hat{L}_\chi,\hat{\sigma}\Big]_+.
\end{split}
\end{equation}
Note that the simplification above might not be possible for a different cost function $f^0$. Now, the CDJP stochastic Hamiltonian in Eq.~\eqref{H_pontrya_ctsms} becomes
\begin{equation}
\begin{split}
    \mathcal{H}(\hat{\sigma}, \hat{\rho}, r, \chi)=&\expval{i[\hat{H}_\chi,\hat{\sigma}]-\frac{1}{4\tau}\left[\hat{V}_\chi,\hat{\sigma}\right]_+}\\&-\frac{1}{2\tau}(r-r_\chi)^2+\frac{1}{2\tau}r_\chi^2,
\end{split}
\label{H_pontrya_ctsms_1}
\end{equation}
with
\begin{equation}
    r_\chi =\expval{\hat{L}_\chi}+ \frac{1}{2}\expval{\Big[\Delta\hat{L}_\chi,\hat{\sigma}\Big]_+}.
\end{equation}
For the most likely readout, according to Pontryagin's maximum principle, the Hamiltonian achieves supremum for given $\hat{\sigma}, \hat{\rho}$ and $\chi$. Thus, $r=r_\chi$ is the most likely readout. Again, here we mathematically treat the readouts $r$ on an equal footing with the controls $\chi$. In case there is a bound on the readouts due to hardware constraints, we need to take it into account for finding the supremum of Eq.~\eqref{H_pontrya_ctsms_1}. However, the CDJP Hamiltonian in Eq.~\eqref{H_pontrya_ctsms_1} is a parabolic function of the readouts. Therefore, the global supremum with respect to the readouts is easy to find and will be adopted as the optimal readout throughout our analysis.

As mentioned after Eq.~\eqref{opt_readout}, the first part in the optimal readout corresponds to the signal, and the second part denotes the optimal noise. In general, the optimal readout has noise due to the presence of detector shot noise in continuous measurements \cite{PhysRevA.47.642}. However, one might design feedback protocols to cancel out the noise \cite{karmakar2025noisecancelingquantumfeedbacknonhermitian}. Additionally, if the optimization involves free endpoints,  Eqs.~\eqref{PMP_cond_trans_0}  and \eqref{P_cond_trans_f} suggest we can choose $\hat{\sigma}=0$, leading to zero optimal noise.

The optimal Hamiltonian can be expressed as 
\begin{equation}
\begin{split}
    \mathcal{H}^\star(\hat{\sigma}, \hat{\rho},  \chi)=&\expval{i[\hat{H}_\chi,\hat{\sigma}]-\frac{1}{4\tau}\left[\hat{V}_\chi,\hat{\sigma}\right]_+}+\frac{1}{2\tau}r_\chi^2,
\end{split}
\label{H_pontrya_ctsms_2}
\end{equation}
which achieves supremum for the optimal control $\chi$, for given $\hat{\sigma}$ and $\hat{\rho}$. Note that the form of the above Hamiltonian is the same as Eq.~\eqref{h_op_general}, as expected.

Thus, when the cost function is the CDJ action, Pontryagin's maximum principle gives the most likely path equations. Also, as explained in the Appendix \ref{CDJ_vs_costate}, the CDJ auxiliary variables can be thought of as a particular parametrization of the costate operator. 

\subsection{\label{control_SHO}Optimal control for a monitored harmonic oscillator}

We apply the preceding formalism to find optimal control protocols for a monitored harmonic oscillator. First, we introduce a parametric control to the harmonic oscillator Hamiltonian
\begin{equation}
    \hat{H}=\frac{1}{2}\left(\hat{X}^2+\hat{P}^2\right)+\lambda_1(t)\hat{X}^2,
    \label{text_parametricH}
\end{equation}
where $\chi_1(t)\equiv \lambda_1(t)$ is a control parameter with its value bounded by $\lambda_{1}^{\max}$. The additional $\hat{X}^2$ term can represent oscillators with tunable frequencies. Such parametric Hamiltonians have been previously considered for cooling \cite{PhysRevA.107.023516, PhysRevA.107.053521}. The oscillator is undergoing adaptive quadrature measurements expressed by the observable $\hat{L}_\theta$ in Eq.~\eqref{l_theta}. We consider all the trajectories starting from a specific initial state $\hat{\rho}(t=0)=\hat{\rho}_i$ and reaching some specified final state $\hat{\rho}(t=t_f)=\hat{\rho}_f$. We ask, what choice of the controls $\lambda_1$ and $\theta$, as functions of time, will maximize the readout probabilities? In other words, we look for the trajectory with the highest probability under the allowed control values. Note, our formalism also applies to problems involving just coherent controls or just dissipative controls. In the absence of measurements, the cost functional Eq.~\eqref{gen_cost_fn} has to be defined accordingly. We do not claim that the state preparation examples presented here provide the best strategy to prepare these specific target states. They provide the optimal controls to maximize the readout probabilities in the presence of the parametric potential and quadrature measurements. In other words, they demonstrate the efficacy of the costate-based general CDJP formalism.

The conditional state evolution along the most likely path is expressed by Eq.~\eqref{op_rho_sho}, with the system  Hamiltonian given by Eq.~\eqref{text_parametricH}. To look at the most likely readout evolution, we define the following scalars (see Appendix \ref{app_parametric})
\begin{equation}
    \begin{split}
\Gamma(n,m)&=\textrm{Tr}\left(\hat{X}^n\hat{P}^m\hat{\Omega}\right),\\\kappa(n,m)&=i\textrm{Tr}\left(\hat{X}^n\hat{P}^m\hat{\Lambda}\right),
    \end{split}
    \label{Gamma_Kappa_defn_txt}
\end{equation}
with $\hat{\Omega}=\frac{1}{2}\left[\hat{\rho},\hat{\sigma}\right]_+$ and $\hat{\Lambda}=\left[\hat{\rho},\hat{\sigma}\right]$. Then, the optimal readout is given by
\begin{equation}
    r_\theta = \cos\theta\Gamma(1,0)+\sin\theta\Gamma(0,1).
\end{equation}
We can find the optimal readout evolution  by integrating the following set of equations
\begin{subequations}
     \begin{align}
         \begin{split}
             \frac{d\Gamma(1,0)}{dt}=&\Gamma(0,1) \\&-
             \frac{\sin\theta}{4\tau}\left(\cos\theta\kappa(1,0)+\sin\theta\kappa(0,1)\right),
         \end{split}\\
          \begin{split}
             \frac{d\Gamma(0,1)}{dt}=&-(1+2\lambda_1)\Gamma(1,0)\\&+
             \frac{\cos\theta}{4\tau}\left(\cos\theta\kappa(1,0)+\sin\theta\kappa(0,1)\right),
         \end{split}\\
         \begin{split}
             \frac{d\kappa(1,0)}{dt}=&\kappa(0,1), \quad  \frac{d\kappa(0,1)}{dt}=-\tilde{\lambda}_1\kappa(1,0),
         \end{split}
     \end{align}
     \label{first_G_k_txt}
 \end{subequations} 
where $\tilde{\lambda}_1=(1+2\lambda_1)$. The optimal Hamiltonian in Eq.~\eqref{H_pontrya_ctsms_2} can be written as
\begin{equation}
    \begin{split}
&\mathcal{H}^\star(\hat{\sigma},\hat{\rho},\theta,\lambda_1,\lambda_2) =\\& -i\textrm{Tr}\left(\hat{H}\hat{\Lambda}\right)+\frac{1}{2\tau}\left(\textrm{Tr}\left(\hat{L}_\theta\hat{\Omega}\right)\right)^2-\frac{1}{2\tau}\textrm{Tr}\left(\hat{L}_\theta^2\hat{\Omega}\right),
    \end{split}
\end{equation}
which can be expressed in terms of the Eqs.~\eqref{Gamma_Kappa_defn_txt} as
\begin{equation}
\begin{split}
        \mathcal{H}^\star&=-\Big(\frac{1}{2}\tilde{\lambda}_1\kappa(2,0)+\frac{1}{2}\kappa(0,2)\Big)\\&+\frac{1}{2\tau}\Big(A_\Gamma\cos 2\theta+B_\Gamma\sin 2\theta\Big)\\&+\frac{1}{4\tau}\Big(\Gamma(1,0)^2+\Gamma(0,1)^2-\Gamma(2,0)-\Gamma(0,2)\Big),
        \label{hstar_1_txt}
\end{split}
\end{equation}
where 
\begin{equation}
    \begin{split}
        A_\Gamma&=\frac{1}{2}\Big(\Gamma(1,0)^2-\Gamma(0,1)^2-\Gamma(2,0)+\Gamma(0,2)\Big)\\B_\Gamma&=\Gamma(1,0)\Gamma(0,1)-\tilde{\Gamma}(1,1).
    \end{split}
\end{equation}
We define $\tilde{\Gamma}(1,1)=\Gamma(1,1)-\frac{i}{2}$ such that $B_\Gamma$ is real. Then, according to Condition I of Pontryagin's maximum principle, the optimal value of $\lambda_1$ is given by
\begin{equation}
    \lambda_1^{\star}(t)=-\lambda_1^{\max}\textrm{sign}\left(\kappa(2,0)\right).
    \label{l1_opt_txt}
\end{equation}
Thus, the Pontryagin maximum principle analytically shows that the optimal parametric potential should have a bang-bang form.   Similarly, we define $A_\Gamma = R_\Gamma\cos\phi_\Gamma$ and $B_\Gamma = R_\Gamma\sin\phi_\Gamma$ such that $R_\Gamma=\sqrt{A_\Gamma^2+B_\Gamma^2}\geq0$. Then the final term in Eq.~\eqref{hstar_1_txt} can be written as $\frac{R_\Gamma}{2\tau}\cos(\phi_\Gamma-2\theta)$, which should take its maximum value for optimal $\theta$. Assuming $\phi_\Gamma\in [-\pi,\pi]$ and $\theta\in [-\tfrac{\pi}{2},\tfrac{\pi}{2}]$ (since replacing $\hat{L}_\theta$ by $-\hat{L}_\theta$ does not change the optimal paths),  the optimal $\theta$ is given by
\begin{equation}
    \theta^{\star}(t)=\frac{\phi_\Gamma}{2}.
    \label{theta_opt_txt}
\end{equation}
For optimal control, the value of the optimal Hamiltonian is given by
\begin{equation}
\begin{split}
&\mathcal{H}^\star=\\&\lambda_1^{\max}|\kappa(2,0)|-\frac{1}{2}\Big(\kappa(2,0)+\kappa(0,2)\Big) +\frac{1}{2\tau}R_\Gamma\\&+\frac{1}{4\tau}\Big(\Gamma(1,0)^2+\Gamma(0,1)^2-\Gamma(2,0)-\Gamma(0,2)\Big).
\end{split}
        \label{hstar_3_txt}
\end{equation}
The evolutions of the  second order $\Gamma$ terms are given by  \begin{subequations}
     \begin{align}
         \begin{split}
             \frac{d\Gamma(2,0)}{dt}=&2\tilde{\Gamma}(1,1) +
             \frac{\sin\theta}{2\tau}\Big(r_\theta\kappa(1,0)\\&-\cos\theta\kappa(2,0)-\sin\theta\kappa(1,1)\Big),
         \end{split}\\
          \begin{split}
             \frac{d\tilde{\Gamma}(1,1)}{dt}=&-\tilde{\lambda}_1\Gamma(2,0)+\Gamma(0,2)\\&
             +\frac{1}{4\tau}\Bigg(r_\theta\Big(\sin\theta\kappa(0,1)-\cos\theta\kappa(1,0)\Big)\\&+\Big(\cos^2\theta\kappa(2,0)-\sin^2\theta\kappa(0,2)\Big)\Bigg),
         \end{split}\\ \begin{split}
             \frac{d\Gamma(0,2)}{dt}=&-2\tilde{\lambda}_1\tilde{\Gamma}(1,1)+\frac{\cos\theta}{2\tau}\Bigg(-r_\theta\kappa(0,1)\\&
             +(\sin\theta\kappa(0,2)+\cos\theta\kappa(1,1)\Big)\Bigg).
         \end{split}
         \end{align}
       \label{secondorder_G_txt}
       \end{subequations}Similarly, the second order $\kappa$ evolutions are given by.
       \begin{subequations}
       \begin{align}
         \begin{split}
             \frac{d\kappa(2,0)}{dt}=&2\kappa(1,1) +
             \frac{2\sin\theta}{\tau}\Bigg(-r_\theta\Gamma(1,0)+\\&\cos\theta\Gamma(2,0)+\sin\theta\tilde{\Gamma}(1,1)\Bigg),
         \end{split}\\
          \begin{split}
             \frac{d\kappa(1,1)}{dt}=&-\tilde{\lambda}_1\kappa(2,0)+\kappa(0,2)\\&+
             \frac{1}{\tau}\Bigg(r_\theta\Big(\cos\theta\Gamma(1,0)-\sin\theta\Gamma(0,1)\Big)\\&+\Big(-\cos^2\theta\Gamma(2,0)+\sin^2\theta\Gamma(0,2)\Big)\Bigg),
         \end{split}\\ \begin{split}
             \frac{d\kappa(0,2)}{dt}=&-2\tilde{\lambda}_1\kappa(1,1)+\frac{2\cos\theta}{\tau}\Bigg(r_\theta\Gamma(0,1)\\&
             -\Big(\sin\theta\Gamma(0,2)+\cos\theta\tilde{\Gamma}(1,1)\Big)\Bigg).
         \end{split}
     \end{align}
     \label{secondorder_k_txt}
 \end{subequations}Note that the equations \eqref{first_G_k_txt}, \eqref{secondorder_G_txt} and \eqref{secondorder_k_txt} are self-contained. Thus, the optimal controls can be found by integrating Eq.~\eqref{op_rho_sho} along with Eqs.~\eqref{first_G_k_txt}, \eqref{secondorder_G_txt}, and \eqref{secondorder_k_txt} with the control values given by Eqs.~\eqref{l1_opt_txt} and \eqref{theta_opt_txt}. Instead of integrating the costate $\hat{\sigma}$, which has the same dimension as $\hat{\rho}$, we just need to integrate ten scalars $
\Gamma(1,0)$, $\Gamma(1,0)$, $\kappa(1,0)$, $\kappa(0,1)$, $\Gamma(2,0)$, $\tilde{\Gamma}(1,1)$, $\Gamma(0,2)$, $\kappa(2,0)$, $\kappa(1,1)$, $\kappa(0,2)$. 

Pontryagin's maximum principle provides a necessary condition on the optimality of the protocol. Thus, after solving equations \eqref{first_G_k_txt}, \eqref{secondorder_G_txt}, and \eqref{secondorder_k_txt} with optimal controls Eqs.~\eqref{l1_opt_txt} and \eqref{theta_opt_txt}, we need to ensure the cost function \eqref{Jcost} obtains the minimum possible value. Exploring control protocols for state preparation that guarantee a unique solution will be a worthwhile endeavor.

In the following subsections, we will apply our results to find optimal controls for state preparation. The CDJP framework maximizes the probabilities of readouts for given initial and final states. The optimal controls found in this framework are expected to facilitate trajectories with favorable outcomes.  In general, to find optimal controls for state preparation, one needs to consider free endpoint problems   (see Ref.~\cite{zhang2025optimalschedulemultichannelquantum}). For a given initial state and free final state, one needs to consider the final endpoint transversality condition in Eq.~\eqref{P_cond_trans_f}. However, such an analysis is beyond the scope of this article. Moreover, we will show that, even under a fixed-endpoint assumption, a significant fraction of trajectories can be brought very close to the target state.  One can also incorporate the readout dependence of the controls, as would be appropriate in scenarios with continuous feedback \cite{PhysRevA.49.2133}. The general formalism presented in Sec.~\ref{text_pontrya} allows a straightforward adaptation of the most likely path based optimization for stochastic master equations with feedback. Also, in the following examples, we assume ideal detectors with no extra decoherence for simplicity.

\subsection{\label{examples_PMP_1}Example 1: Preparation of a Binomial Code Word}
\begin{figure*}[t]
     \centering
    \includegraphics[width = .95\textwidth]{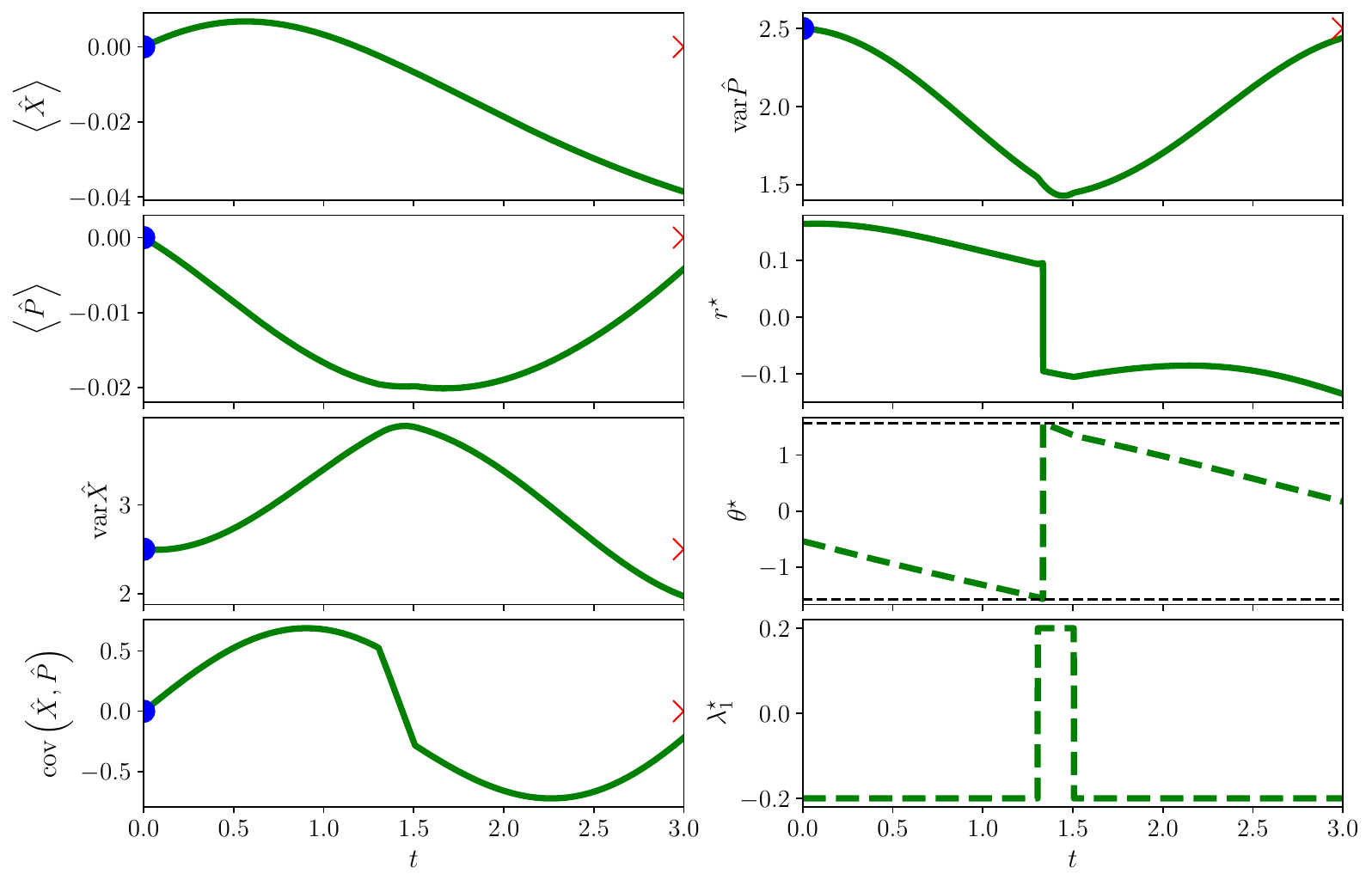}
    \caption{Optimal control for $\ket{\psi(t=0)}=\ket{\psi_i}=\frac{\ket{0}-\ket{4}}{\sqrt{2}}$ and $\ket{\psi(t=3.0)}=\ket{\psi_f}=\frac{\ket{0}+\ket{4}}{\sqrt{2}}$ with collapse timescale $\tau=15.0$. Time and the collapse timescale are in units of the inverse of the oscillator frequency. The left panels from top to bottom show the time evolution of the position and momentum expectation values, position variance, and the covariance of position and momentum, respectively, under optimal control.  The top panel on the right-hand side shows the evolution of the momentum variance. All the observables are scaled to be dimensionless (see Sec.~\ref{sho}). The blue dots show the initial state, and the red `$\times$' show the final state. The second panel (from the top) on the right-hand side shows the most likely readout under optimal control. The third and fourth panels on the right-hand side show the optimal control parameters $\theta_1^\star$ and $\lambda_1^\star$. In other words, they show the optimal measurement quadrature and parametric potential strength. As explained in the text, the potential strength,    $\lambda_1^{\star}$, is of ``bang-bang'' form. The final state here is reached with 95.46\% fidelity (see Appendix \ref{numerics_OC}).}
    \label{fig:optimal_control_binomial_code}
\end{figure*}
\begin{figure}[!htbp]
	\includegraphics[width=0.90\linewidth,trim = {0 0 0 0}, clip]{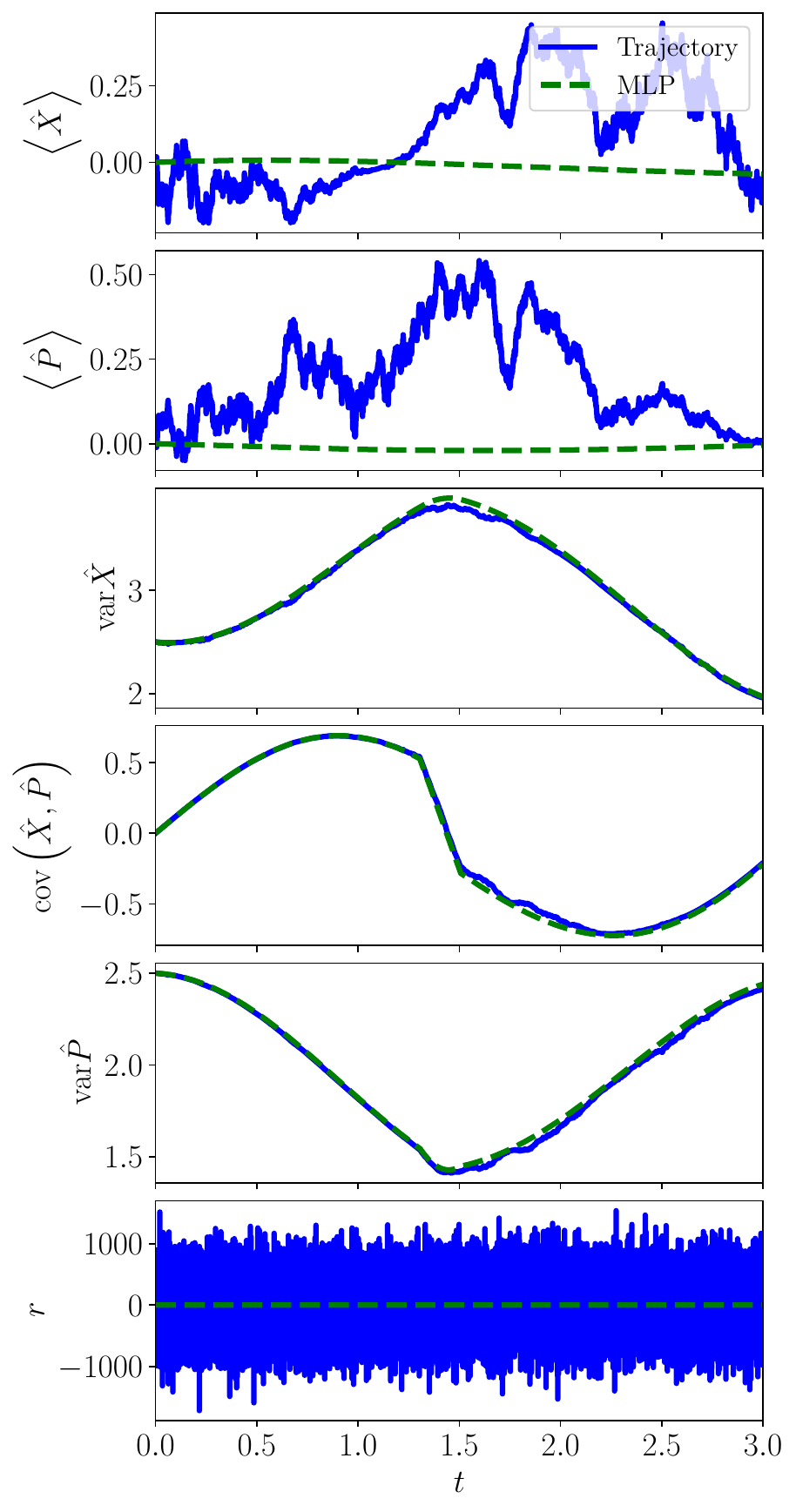}
	\caption{From top to bottom, the panels show the time evolution of the position expectation value, momentum expectation value, the variance of position, the covariance of position and momentum, the variance of momentum, and the readout under the control in Figure~\ref{fig:optimal_control_binomial_code}. The green dashed line shows the most likely path under optimal control shown in Figure~\ref{fig:optimal_control_binomial_code}. The blue curve shows a sample simulated stochastic trajectory under the optimal control shown in the bottom two right panels of Figure~\ref{fig:optimal_control_binomial_code}, starting from the initial state $\ket{\psi_i}=\frac{\ket{0}-\ket{4}}{\sqrt{2}}$. The collapse timescale is $\tau=15.0$ (time and the collapse timescale are in units of the inverse of the oscillator frequency). The quadratures and the readout are dimensionless (see Sec.~\ref{sho}).  We see that the expectation values along the trajectory jitter around the most likely path.  Additionally, the trajectory closely follows the evolution of the variances and the covariance. The expectation values for the most likely path and the optimal readout appear flat because the corresponding scales involved with the stochastic trajectory are much larger (see Appendix \ref{numerics_trajectory}  for the numerical method adopted to simulate trajectories).}
	\label{trajector_binomial_code}%
\end{figure}

Now, we apply our general results to problems in Bosonic quantum computing. In the subsequent analysis, the final time is $t_f=3.0$ and the collapse time scale $\tau=15.0$, both in units of the inverse of the harmonic oscillator frequency. The quadratures and the readouts are made dimensionless (see Sec.~\ref{sho}).

Our first example pertains to Binomial codes, which represent logical qubits in terms of coherent superpositions of the Fock states of a harmonic oscillator \cite{PhysRevX.6.031006, 10.21468/SciPostPhysLectNotes.70}. Depending on the number of Fock states used, such codes can protect against single/multiple photon loss/gain errors and dephasing errors. We look at the simplest example of a binomial code, which defines the logical code words as $\ket{0_\textrm{L}}=\frac{\ket{0}+\ket{4}}{\sqrt{2}}$ and $\ket{1_\textrm{L}}=\ket{2}$, respectively. Such code can protect against a single photon loss. If a photon jump occurs, the code words transform to $\ket{E_1} = \ket{3}$ and  $\ket{E_2} = \ket{1}$, respectively. However, if no photon jump is observed, the $\ket{0_\textrm{L}}$ state can still undergo error within a subspace defined by $\ket{0_\textrm{L}}$ and $\ket{E_0}=\frac{\ket{0}-\ket{4}}{\sqrt{2}}$. Thus, the state $\ket{E_0}$ corresponds to an error word under no jump evolution \cite{PhysRevX.6.031006, 10.21468/SciPostPhysLectNotes.70}.  We consider the evolution from the initial state $\ket{\psi(t=0)}=\ket{\psi_i}=\ket{E_0}=\frac{\ket{0}-\ket{4}}{\sqrt{2}}$ to the final state $\ket{\psi(t=3.0)}=\ket{\psi_f}=\ket{0_\textrm{L}}=\frac{\ket{0}+\ket{4}}{\sqrt{2}}$ with measurement collapse timescale $\tau=15.0$ and $\lambda_1^{\max}=0.2$.  We wish to solve for the adaptive quadrature measurement and parametric Hamiltonian control that minimizes Eq.~\eqref{Jcost} amongst the trajectories starting from $\ket{\psi_i}$ and reaching $\ket{\psi_f}$. The results of the optimization are shown in Figure~\ref{fig:optimal_control_binomial_code}. Here, the final state is reached with 95.46\% fidelity (see numerical methods in Appendix \ref{numerics_OC}). The figure shows the most likely path under optimal control in the four left and topmost right panels. The second panel on the right-hand side shows the most likely readout under optimal control. The optimal measurement quadrature and parametric potential strengths are shown in the bottom two panels on the right. As per Eq.~\eqref{l1_opt_txt}, the optimal parametric potential is of the ``bang-bang'' form with two sign switches close to $t=1.5$.  The optimal quadrature shows a saw-tooth-like evolution with a discontinuity near $t=1.4$.

In Figure~\ref{trajector_binomial_code}, we compare the evolutions of a sample stochastic trajectory (blue solid) and the most likely path (green dashed) under the optimal control. We see that the trajectory follows the most likely path very closely.  Note, for Gaussian states undergoing continuous quadrature measurements, the covariance matrix elements evolve deterministically \cite{PRXQuantum.3.010327}.   However, the initial state corresponding to Figure~\ref{trajector_binomial_code} is non-Gaussian. In such a scenario, the evolution of the variances and the covariance is stochastic under continuous measurements, consistent with our findings in Figure~\ref{trajector_binomial_code}.

\begin{figure}[!t]
\begin{tikzpicture}
    \node[] at (0,0) {\includegraphics[width = .9\linewidth]{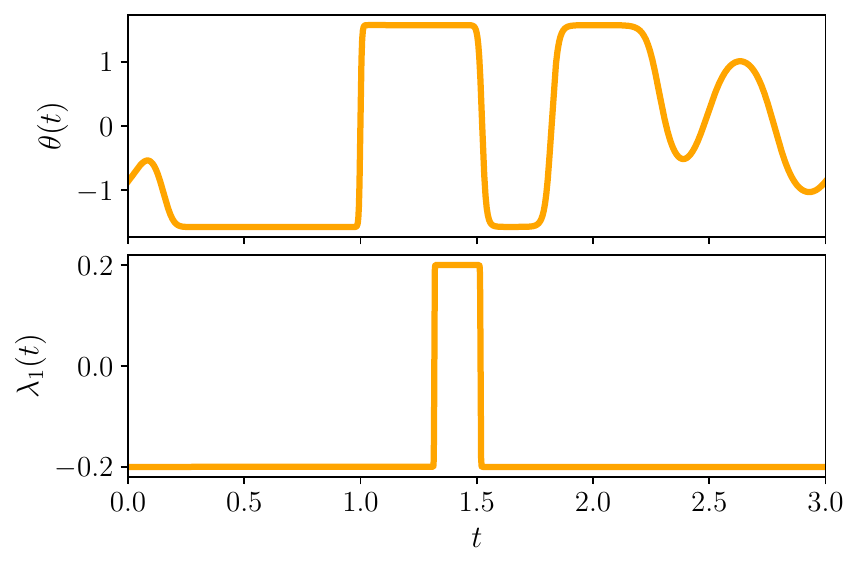}};
     \node[] at (0,-5) {\includegraphics[width = 0.9\linewidth]{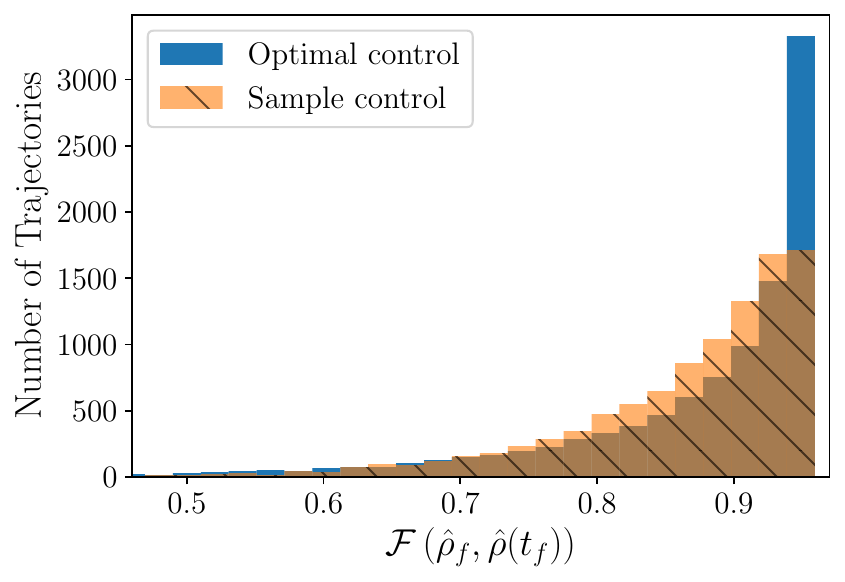}};
	\node[] at (-3.5,2.3) {(a)};
	\node[] at (-3.5,-2.4) {(b)};
	\end{tikzpicture}
    \caption{In panel (a), the top (bottom) plot shows sample measurement quadrature (parametric potential strength) as a function of time for collapse timescales $\tau=15.0$ (in units of $1/$oscillator frequency)  such that trajectories starting from $\ket{\psi_i}=\frac{\ket{0}-\ket{4}}{\sqrt{2}}$ have a significant probability of reaching $\ket{\psi_f}=\frac{\ket{0}+\ket{4}}{\sqrt{2}}$ at $t=3.0$. Note, $\lambda_1(t)$ has a structure almost identical to $\lambda_1^\star(t)$ shown in Figure~\ref{fig:optimal_control_binomial_code}. However, $\theta(t)$ here is very different  from $\theta^\star(t)$.  Panel (b) shows the histogram of final state fidelities (with respect to the target state $\ket{\psi_f}=\frac{\ket{0}+\ket{4}}{\sqrt{2}}$) of 10,000 simulated trajectories under the optimal control in Figure~\ref{fig:optimal_control_binomial_code} (blue, vertical hatched) and the sample control presented in panel (a) (orange, diagonal hatched). We see that a significantly higher number of trajectories can achieve very high fidelities under optimal control. }\label{fig:sample_control_binomialcode}
\end{figure}

It is insightful to compare the statistics of the trajectories under the optimal control with those under a sample control.  For a given initial state, controls, and final time, we expect the trajectories to be clustered near some final state. However, we are interested in specific initial and final (target) states. Thus, arbitrarily chosen controls will likely not result in a trajectory that reaches close to the chosen final state. To that end, we generate sample controls that overcome this issue, allowing a fair comparison with the performance of the optimal control. Using Fourier series expansion, we find controls $\theta(t)$ and $\lambda_1(t)$ such that there is a most likely path between the initial and the final states (see Appendix \ref{numerics_sample_control} for the relevant numerical methods).   In other words, we are interested in generating a sample control $\theta(t)$ and $\lambda_1(t)$ such that trajectories starting from $\ket{\psi_i}$ at $t=0$ have a non-infinitesimal probability of reaching $\ket{\psi_f}$ at $t=3.0$. Note, controls found in this manner are unlikely to be optimal since they need not satisfy Condition I (Eq.~\eqref{PMP_cond_hmax})  of the PMP. Figure~\ref{fig:sample_control_binomialcode}(a) provides one such example. Here, the final state can be reached with 95.58\% fidelity. Figure~\ref{fig:sample_control_binomialcode}(b) shows the histograms of the final state fidelities (with respect to the target state) for trajectories under optimal control and those under the sample control. Under the optimal control, 57.08\% (18.31\%) of trajectories have fidelities of more than 90\% (95\%). With the sample control, 46.01\% (6.19\%) of the trajectories have a fidelity of more than 90\% (95\%). This is around a 24\% (196\%) increase in the number of trajectories reaching 90\% (95\%) fidelities. Since the optimal control was computed with the assumption of fixed endpoints, the trajectories generated do not correspond to maximizing the fraction of trajectories reaching the target state. Rather, here we maximize the readout probabilities of the trajectories that reach the target state. However,  Figure~\ref{fig:sample_control_binomialcode}\,(b) shows that the control derived in this manner can still be useful for state preparation since a significant portion of the trajectories reach very close to the target state. We note that the histogram in Figure~\ref{fig:sample_control_binomialcode}\,(b) does not constitute a formal proof of the optimality of the protocol. To ensure optimality,  we must restrict ourselves to trajectories between $\ket{\psi_i}$ and $\ket{\psi_f}$ and consider all possible controls.  The proof of the optimality of a specific protocol is beyond the scope of this article. 
\begin{figure*}[t]
    % \centering
    \includegraphics[width = .95\textwidth]{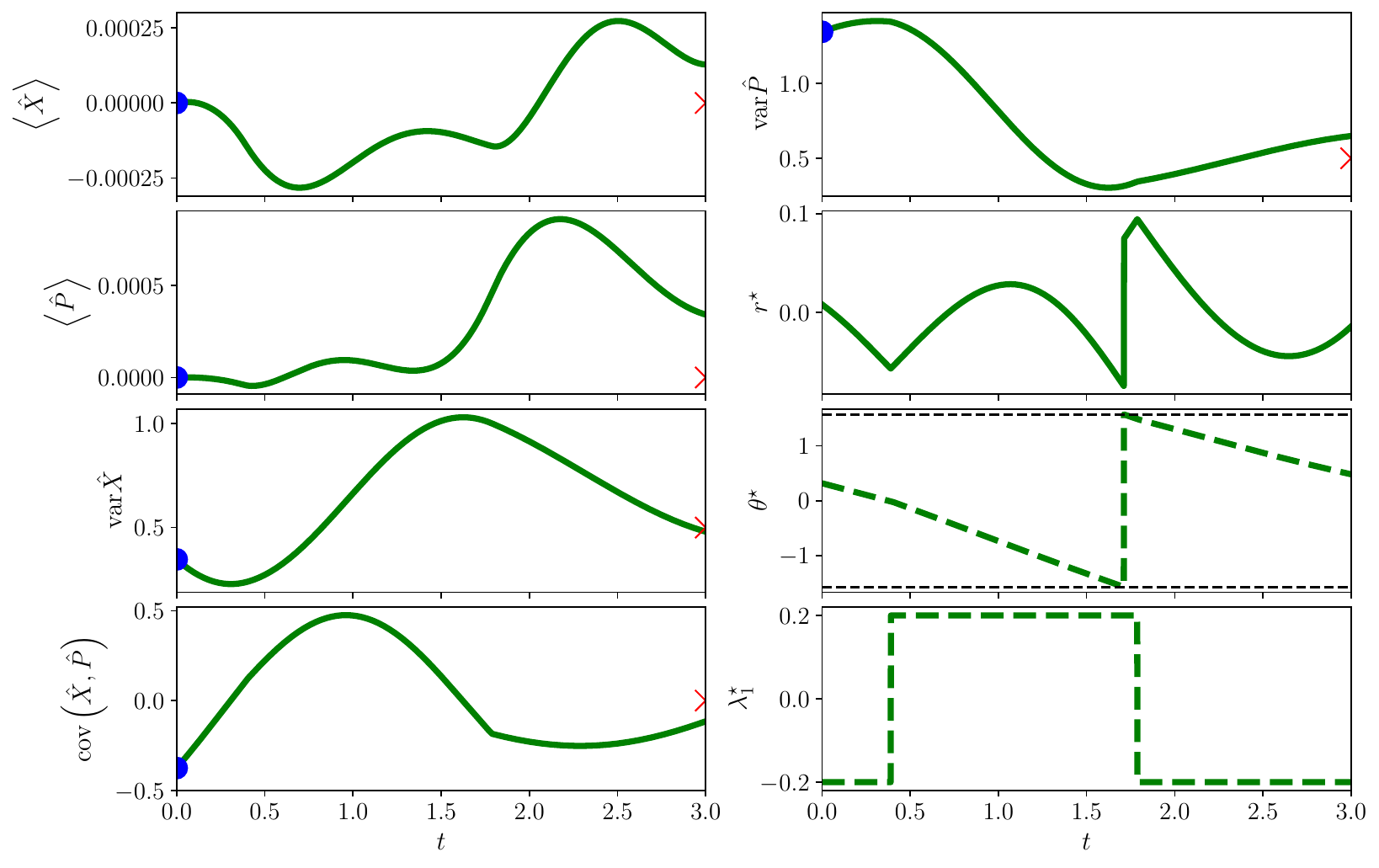}
   \caption{Optimal control for $\ket{\psi(t=0)}\propto\ket{\alpha}+\ket{-\alpha}$ with $\alpha=0.25-i0.75$ and $\ket{\psi(t=3.0)}=\ket{\psi_f}=\ket{0}$. The collapse timescale is $\tau=15.0$ (in units of $1/$oscillator frequency).  Again, the left panels from top to bottom show the time evolution of position and momentum expectation values,  position variance, and covariance of position and momentum, respectively, under optimal control.  The top panel on the right-hand side shows the evolution of the momentum variance. The blue dots show the initial state, and the red `$\times$' show the final state. The second panel (from the top) on the right-hand side shows the most likely readout under the optimal control. The third and fourth panels on the right-hand side show the optimal quadrature $\theta^\star$ and parametric potential strength $\lambda_1^\star$. Again, we see a ``bang-bang'' form for $\lambda_1^\star$  with value $\lambda_1=0.2$ approximately between $0.4<t<1.8$. $\theta^\star$ has a discontinuity near $t\approx 1.7$.  The final state here is reached with 97.49\% fidelity.}
\label{fig:optimal_control_catstate}
\end{figure*}
\subsection{\label{examples_PMP_2}Example 2: Parametric cooling from a cat state}
\begin{figure}[!t]
     \begin{tikzpicture}
    \node[] at (0,0) {\includegraphics[width = .9\linewidth]{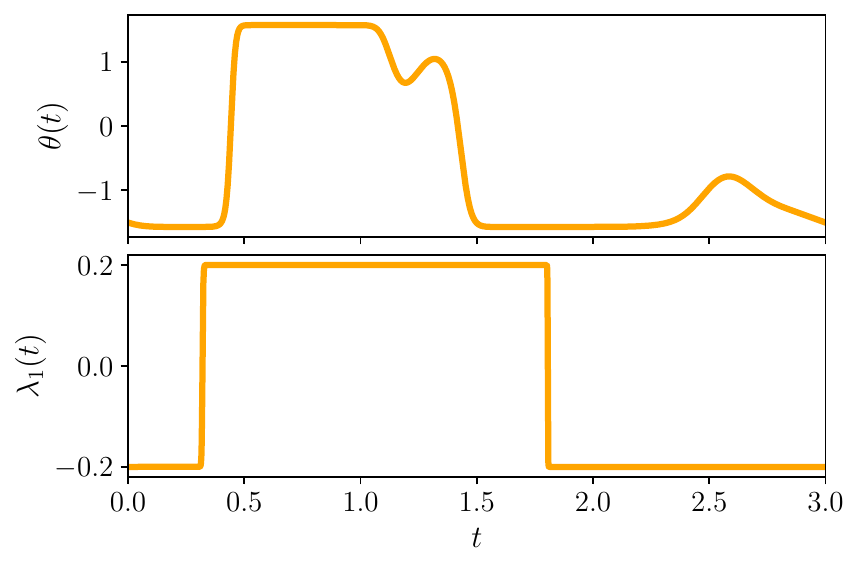}};

     \node[] at (0,-5) {\includegraphics[width = 0.9\linewidth]{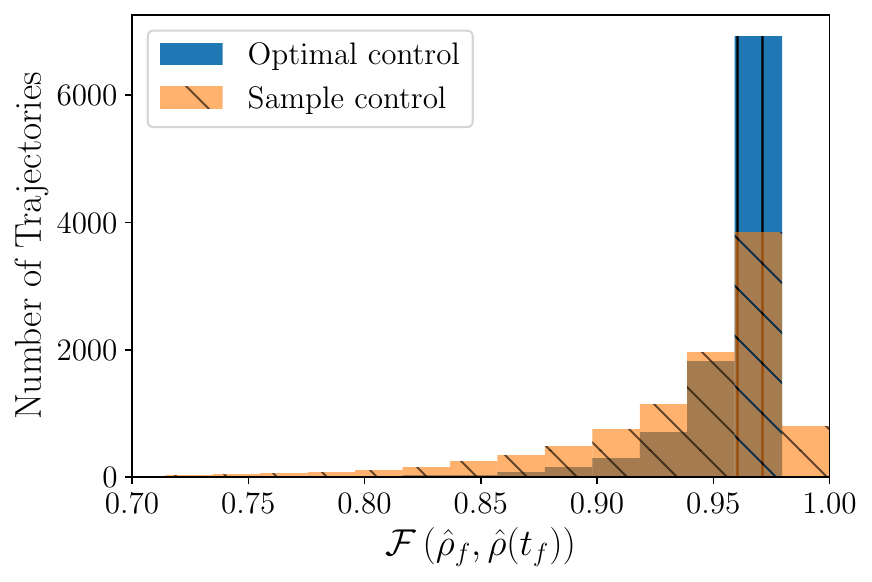}};
	\node[] at (-3.5,2.3) {(a)};
	\node[] at (-3.5,-2.4) {(b)};
	\end{tikzpicture}
    \caption{In panel (a), the top (bottom) plot shows sample measurement quadrature (parametric potential strength) as a function of time such that trajectories starting from $\ket{\psi_i}\propto\ket{\alpha}+\ket{-\alpha}$ with $\alpha=0.25-i0.75$ have a significant probability of reaching $\ket{\psi_f}=\ket{0}$ at $t=3.0$ with $\tau=15.0$. Time and collapse timescales are in units of the inverse of the oscillator frequency. Again, $\lambda_1(t)$ has a structure very similar to the $\lambda_1^\star(t)$ shown in Figure~\ref{fig:optimal_control_catstate}. $\theta(t)$ here is very different  from $\theta^\star(t)$.  Panel (b) shows the histogram of final state fidelities (with respect to the target state $\ket{\psi_f}=\ket{0}$) of 10,000 simulated trajectories under the optimal control in Figure~\ref{fig:optimal_control_catstate} (blue, vertical hatched) and the sample control presented in panel (a) (orange, diagonal hatched). A significantly higher number of trajectories can reach large fidelities than trajectories under the sample control. We observe a small number of trajectories reach very close to fidelity 1.0 under the sample control. This is because the numerical simulation for finding the sample control can achieve better convergence to the target state in this case (see Appendices \ref{numerics_sample_control} and \ref{numerics_OC}).}
    \label{fig:sample_control_catstate}
\end{figure}

Two-component Schr\"odinger's cat states constitute an important family of non-Gaussian harmonic oscillator states. Such states are coherent superpositions of coherent states  \cite{gerry_quantum_1997, PhysRevA.59.2631}. They can be generated by two-photon-driven dissipative processes and provide a Bosonic encoding that can protect against dephasing errors \cite{Mirrahimi_2014}. For our second example, we consider the even cat state $\ket{\psi(t=0)}=\ket{\psi_i}\propto\ket{\alpha}+\ket{-\alpha}$ with $\alpha=0.25-i0.75$ as the initial state. The final state is the ground state $\ket{\psi(t=3.0)}=\ket{0}$ with collapse timescale\footnote{Again, time and the collapse timescale $\tau$ are in units of $1/$oscillator frequency.} $\tau=15.0$ and $\lambda_1^{\max}=0.2$.  In other words, we are looking at parametric cooling starting from the initial state $\ket{\psi_i}$ \cite{PhysRevA.107.023516, PhysRevA.107.053521}. The final state should be free to consider the full generality of parametric cooling. For simplicity, we limit our analysis to fixed-endpoint scenarios. We will see shortly that optimal control computed under such restriction can still be very useful. Figure~\ref{fig:optimal_control_catstate} shows the optimal control protocol for the problem.  In this case, our simulations achieved a final state fidelity of 97.49\%. The parametric potential strength is of ``bang-bang'' form and changes sign twice. The optimal quadrature has a discontinuity near $t=1.7$.

Figure~\ref{fig:sample_control_catstate}\,(a) shows a sample control (reaching final state with 98.21\% fidelity) with a trajectory starting from $\ket{\psi_i}$ and reaching  $\ket{\psi_f}$ at $t=3.0$. Note, the $\lambda_1$ is almost identical to the optimal $\lambda_1^{\textrm{opt}}$ shown in Figure~\ref{fig:optimal_control_catstate}. Figure~\ref{fig:sample_control_catstate}\,(b) compares the histograms of the final state fidelities for 10,000 simulated stochastic trajectories under the sample and the optimal control.   Under the optimal control, 97.00\% (79.51\%) of the trajectories have fidelities of more than 90\% (95\%). With the sample control, 84.31\% (56.73\%) of the trajectories have a fidelity of more than 90\% (95\%). This is around a 15\% (39\%) increase in the number of trajectories reaching 90\% (95\%) fidelities. Therefore, we again see a significant increase in the probability of high-fidelity state preparation under optimal control.  Note, in Figure~\ref{fig:sample_control_catstate}\,(b), the sample control can generate trajectories that reach closer to the target state compared to the trajectories generated from the optimal control. This is due to limitations in our numerical methods. To find the optimal control, we integrate Eqs.~\eqref{secondorder_G_txt}, \eqref{secondorder_k_txt} with optimality constraints presented in Eqs.~ \eqref{l1_opt_txt} \eqref{theta_opt_txt}. Moreover, Pontryagin's maximum principle provides necessary but not sufficient conditions for optimality. Thus, during integration, we need to ensure that truly optimal solutions are obtained (see Appendix \ref{numerics_OC} for the details).  We do not need to consider such issues while finding the sample controls. For the above reasons, the numerical simulations to find sample controls converged to higher target-state fidelities. Investigation into sophisticated numerical schemes to overcome these issues is beyond the scope of this article. 

\begin{figure*}[t]
     \centering
    \includegraphics[width = .95\textwidth]{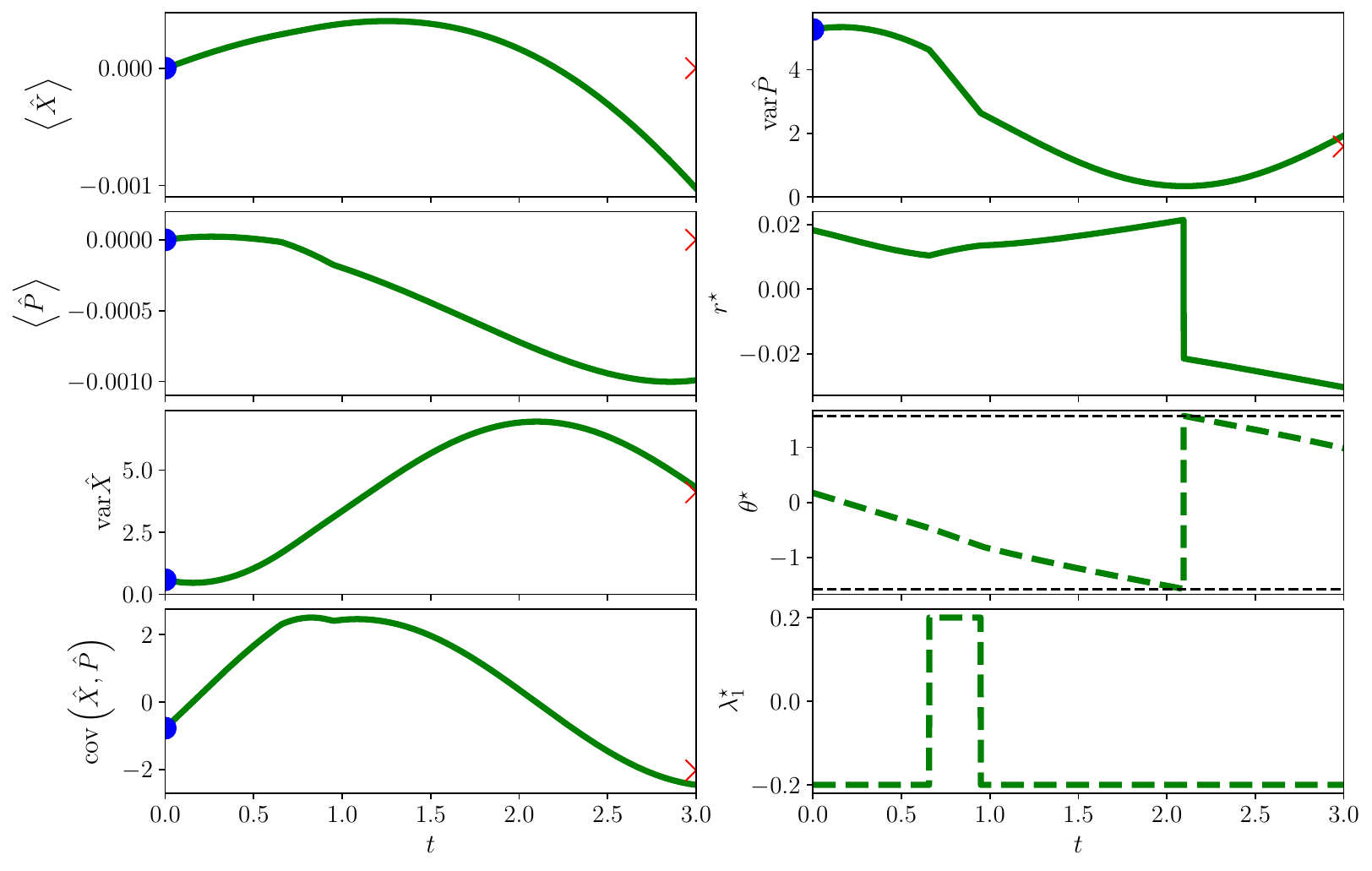}
   \caption{We consider now the cat state to cat state evolution. $\ket{\psi(t=0)}=\ket{\psi_i}\sim\ket{\alpha_i}+\ket{-\alpha_i}$ and $\ket{\psi(t=3.0)}=\ket{\psi_f}\sim\ket{\alpha_f}+\ket{-\alpha_f}$ with $\alpha_i = -0.25+1.55i$ and $\alpha_f = 1.35-0.75i$, and the measurement collapse timescale is $\tau=15.0$.  Time and collapse timescales are in units of the oscillator frequency inverse. The left panels from top to bottom show the time evolution of the expectation values of the position and momentum, and the position variance and covariance of position and momentum, respectively, under optimal control.  The top panel on the right-hand side shows the evolution of the momentum variance. The blue dots show the initial state and the red `$\times$' show the final state. Our numerical method converges to a fidelity of 95.59\%. The second panel (from the top) on the right-hand side shows the most likely readout under the optimal control. The third and fourth panels on the right-hand side show the optimal quadrature $\theta^\star$ and parametric potential strength $\lambda_1^\star$. $\theta^\star$ has a sawtooth-like behavior with a discontinuity near $t=2.1$. $\lambda_1^\star$ has a ``bang-bang'' form with two discontinuities between $0.5<t<1.0$.}
\label{fig:optimal_control_cat2cat}
\end{figure*}
\begin{figure}[!t]
   \begin{tikzpicture}
    \node[] at (0,0) {\includegraphics[width = .9\linewidth]{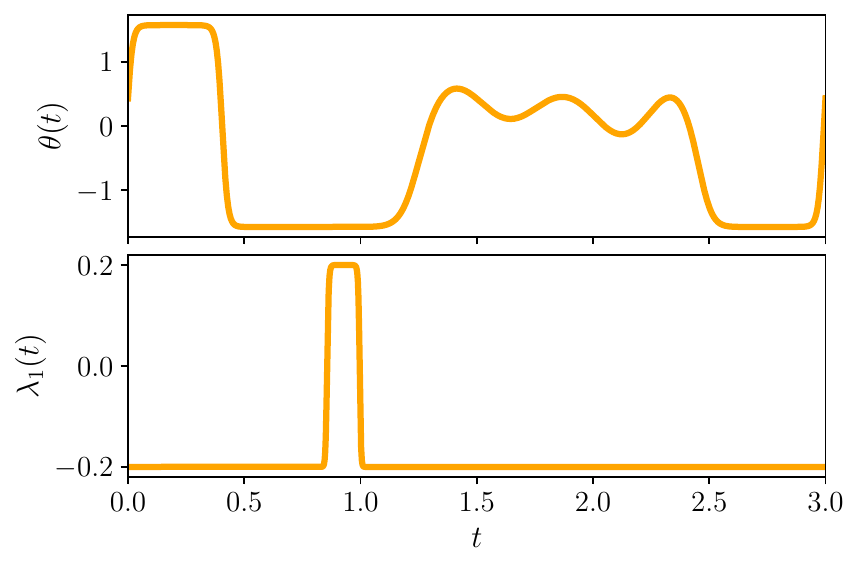}};

     \node[] at (0,-5) {\includegraphics[width = 0.9\linewidth]{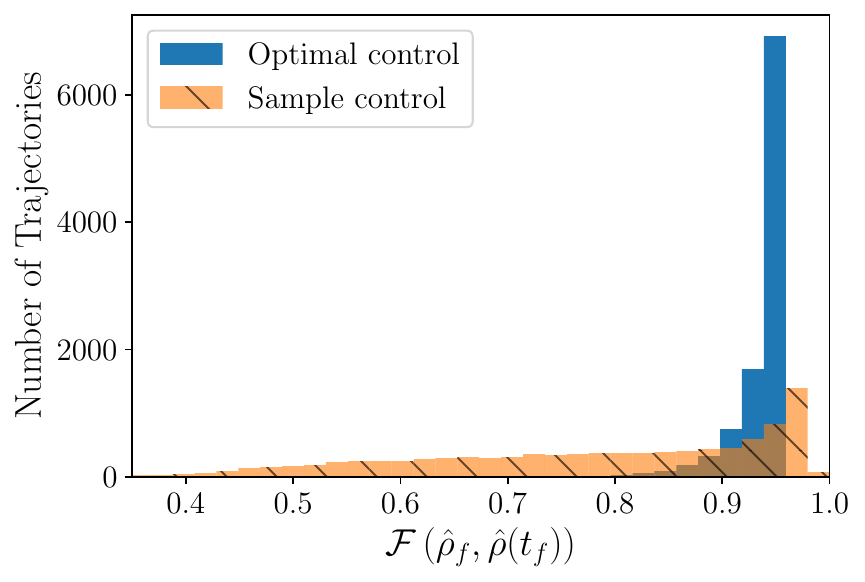}};
	\node[] at (-3.5,2.3) {(a)};
	\node[] at (-3.5,-2.4) {(b)};
	\end{tikzpicture}
    \caption{We consider now the cat state to cat state evolution. $\ket{\psi(t=0)}=\ket{\psi_i}\sim\ket{\alpha_i}+\ket{-\alpha_i}$ and $\ket{\psi(t=3.0)}=\ket{\psi_f}\sim\ket{\alpha_f}+\ket{-\alpha_f}$ with $\alpha_i = -0.25+1.55i$ and $\alpha_f = 1.35-0.75i$ with $\tau=15.0$. Time and collapse timescales are in units of the inverse of the oscillator frequency. Panel (a) shows a sample control that can reach a fidelity of 97.94\%. The parametric potential is similar to the one shown in Figure~\ref{fig:optimal_control_cat2cat}, although the region with $\lambda_1=0.2$ is much thinner. $\theta_1$  is completely different from the optimal quadrature $\theta_1^\star$ shown in Figure~\ref{fig:optimal_control_cat2cat}. Panel (b) shows the histogram of the fidelities of 10,000 trajectories at $t=3.0$ with respect to the target state under optimal control (blue, vertical hatched) and the sample control (orange, diagonal hatched). The optimal control produces high-fidelity states with significantly higher probabilities.}
    \label{fig:sample_control_cat2cat}
\end{figure}
\subsection{\label{examples_PMP_3}Example 3: Cat state to cat state} We now consider a cat state to cat state evolution. $\ket{\psi(t=0)}=\ket{\psi_i}\sim\ket{\alpha_i}+\ket{-\alpha_i}$ and $\ket{\psi(t=3.0)}=\ket{\psi_f}\sim\ket{\alpha_f}+\ket{-\alpha_f}$ with $\alpha_i = -0.25+1.55i$ and $\alpha_f = 1.35-0.75i$. The collapse timescale is $\tau=15.0$ (in units of $1/$oscillator frequency). The optimal control is shown in Figure~\ref{fig:optimal_control_cat2cat}. The final state is reached with 95.59\% fidelity. In this example, $\lambda_1^\star$ has two jumps between $0.5<t<1.0$. The optimal quadrature has a sawtooth behavior with a discontinuity around $t\sim 2.1$.

In Figure~\ref{fig:sample_control_cat2cat}(a), we show a sample control that can reach 97.94\% fidelity. In Figure~\ref{fig:sample_control_cat2cat}\,(b), we compare the histograms of 10,000 trajectories simulated under the optimal control and the sample control.  Applying the optimal control causes more trajectories to reach close to the target state. Under the optimal control, 92.68\% (47.17\%) of trajectories have fidelities of more than 90\% (95\%). With the sample control, 32.78\% (18.89\%) of the trajectories have more than 90\% (95\%) fidelities.  This is around a 182\% (150\%) increase in the number of trajectories reaching 90\% (95\%) fidelities. In this example too, the optimal control demonstrates a superior performance. In this example, we have limited the cat state sizes to $|\alpha|^2\sim 2.4$ due to simulation constraints. The generation of larger cat states and cooling from them in the presence of a non-linear parametric potential (see Appendix \ref{app_parametric}) are interesting future avenues for exploration.

 Again, in the preceding examples, we maximize the readout probabilities of trajectories that start from a specific initial state and reach a fixed final state. However,  these three examples demonstrate that even with a fixed-endpoint analysis, most likely path-based optimal control protocols can generate trajectories that are highly likely to reach a target state.
\section{\label{conclusions}Discussion}In this work, we have formulated a costate-based description of the Chantasri-Dressel-Jordan most likely path formalism for general continuously monitored systems.  We show that the system's evolution along the most likely path can be expressed as a coupled evolution equation of the system's state and a costate operator. Next, we provide a general Pontryagin maximum principle formulation for the optimal control of a quantum system undergoing arbitrary evolution. The cost function is assumed to be arbitrary. For general non-linear evolution, the costate evolution might differ from the negative adjoint evolution prevalent in the literature \cite{Koch_2016, PhysRevApplied.16.054023}. Such differences can be crucial for a continuously monitored system since the conditional dynamics is nonlinear in this case. We derive the CDJ most likely paths as a special case of the general Pontryagin principle when the cost function corresponds to the probability density of the readouts. For this special case, the costate evolution can be reduced to the negative adjoint of the state evolution under an operator transformation.

 The costate-based formulation provides several advantages over the existing auxiliary-variable-based CDJ formalism. The existing auxiliary-variable-based approach requires parametrization of the stochastic master equation for each problem. Afterwards, constructing the stochastic Hamiltonian involves the construction of a stochastic action which quantifies the probability density of the trajectories \cite{PhysRevA.92.032125,Areeya_Thesis}. Such parametrization might become daunting for large systems. One of the primary reasons that the CDJ formalism has not been applied to monitored many-body systems yet is the complexity of parametrization.    Furthermore, it is not clear how one can incorporate additional cost functions apart from the log probability of readouts into the optimization problem. The costate-based approach we presented overcomes the first issue by expressing the most likely paths in terms of the density matrix and the costate operator evolutions. Therefore, standard approaches for simulating many-body dynamics can be adopted to describe the CDJ most likely paths. Also, through the general Pontryagin's maximum principle, one can incorporate several cost functions for optimization. Recently, the costate-based control optimization has been utilized for solving $k$-SAT problems with measurement-driven algorithms \cite{zhang2025optimalschedulemultichannelquantum}.

 We describe a continuously monitored harmonic oscillator with a parametric control potential and undergoing variable quadrature measurements. In this case, we can reduce the costate evolution to 10 scalar parameter evolutions. Such dimensionality reduction facilitates fast numerical simulations. Pontryagin's maximum principle lets us derive the analytical forms of the optimal parametric potential strength and the optimal measurement quadrature. The former is of the ``bang-bang'' form. We look at three fixed-endpoint problems. In the first one, we look at the preparation of binomial code word $\frac{\ket{0}+\ket{4}}{\sqrt{2}}$ starting from $\frac{\ket{0}-\ket{4}}{\sqrt{2}}$. The second example looks at cooling to the ground state starting from an even cat state. In the third example, we examine a cat state to cat state evolution.  For all three examples, we compare the performance of simulated trajectories under the optimal control and a sample control.  We see that trajectories under the optimal control perform significantly better than those under the sample control. Compared to the latter case, under optimal control, a significantly larger number of trajectories can reach high fidelities with respect to the target state.  These findings are consistent with Ref.~\cite{kokaew_quantum_2026}, where most-likely-path-based optimal control protocols were shown to achieve higher success rates in reaching the target state than the Lindbladian mean-path dynamics.  Our results demonstrate that even under the fixed-endpoint assumption, optimal control derived from the CDJ-Pontryagin approach can be very useful for state preparation in continuously monitored systems.

The $\hat{X}^2$ parametric potential and the quadrature measurements offer limited flexibility for state preparation in continuous variable systems. For universal quantum computation, anharmonic potential should be included  \cite{RevModPhys.77.513,PhysRevLett.102.120501}. The limitation of a quadratic potential is evident in our optimal control computations. In simulations, we find the initial values of $\Gamma(n,m)$ and $\kappa(n,m)$ such that their evolution according to Eqs.~\eqref{first_G_k_txt},\eqref{secondorder_G_txt}, \eqref{secondorder_k_txt}, \eqref{l1_opt_txt} and  \eqref{theta_opt_txt} can bring the final state as close as possible to the target state. Our simulations converge around 95-97\% fidelities, while solutions with more than 99\% fidelities would be ideal. Introducing cubic or higher order potential might lead to better final state fidelities. Also, note that Condition I, presented in Sec.~\ref{text_pontrya}, is a necessary condition,  not a sufficient one. In our simulations, we observe multiple possible solutions for the equations presented in Sec.~\ref{control_SHO}. Additional tests are needed to ascertain whether the solution computed from Pontryagin's maximum principle is truly optimal. Our numerical scheme ensures (see Appendix \ref{numerics_OC}) that the cost Eq.~\eqref{Jcost} associated with the optimal control is as small as possible. As the final state fidelity reaches 90-95\%, subsequent iterations of the numerical optimization update the candidate solution only if the cost in Eq.~\eqref{Jcost} is reduced. This leads to a Pareto front optimization between final state fidelity and readout probabilities. Robust numerical methods that guarantee high-fidelity optimal solutions should be investigated in the future.  We defer the formal proof of the optimality of the solutions we found to future explorations.

On a related note, we have previously proven the existence of a unique optimal path for Gaussian state quantum harmonic oscillators \cite{PRXQuantum.3.010327}. The situation might be different for other continuously monitored systems.  It is known that monitored qubits might show multiple optimal paths \cite{LewalleMultipath, NaghilooCaustic, LewalleChaos}. It is worth exploring whether multiple optimal paths can arise for monitored oscillators in non-Gaussian states. Also,  in certain scenarios, optimal paths or an optimal control protocol might not exist. These issues connect broadly to the question of controllability. A thorough exploration of the existence and uniqueness of the optimal paths and the optimal control protocols is needed.

Our work can be useful for finding optimal control for state preparation \cite{PhysRevLett.85.1594,  guerlin_progressive_2007},   dissipative stabilization \cite{harrington_engineered_2022, PhysRevA.107.022216, lewalle2023optimal}, feedback cooling \cite{rossi_measurement-based_2018, Zhang2017, doi:10.1126/science.abh2634}, etc.  Note, the stochastic master equation under feedback is a special case of the general nonlinear evolution presented in Eqs.~\eqref{gen_rho_evol}  and \eqref{eq_F_variation} \cite{PhysRevA.49.2133}.  Thus, the presented formalism can be used to describe continuous feedback, where the applied feedback depends on the readouts. Implementing our framework for finding optimal controls for continuous error correction of Bosonic codes is a natural next step \cite{PhysRevA.67.052310,RevModPhys.77.513, PhysRevX.6.031006, mohseninia_always-quantum_2020, livingston_experimental_2022}. A similar approach can be adopted for autonomous quantum error correction \cite{PhysRevA.90.062344, lebreuilly2021autonomous, gertler_protecting_2021, xu_autonomous_2023}. 

In our work, we adopted the readout probabilities in a continuous measurement as the cost function. Other cost functions, such as the expectation value of the Hamiltonian, might be useful for running quantum algorithms \cite{grange_introduction_2024, PhysRevX.7.021027} or solving classical problems \cite{RevModPhys.94.015004,zhang2024solvingksatproblemsgeneralized}. One can adopt the entanglement of formation as a cost function to prepare entangled states \cite{PhysRevLett.80.2245,PhysRevA.102.062418,Lewalle:21, PhysRevX.6.041052, shea2024stochastic}. With our framework, quantum optimal control for many-body systems can be studied. One example is a dissipative Bosonic Kitaev chain in the presence of quadratic bosonic Lindbladians  \cite{ PhysRevLett.127.245701, PhysRevB.108.214312}. For this system, a Gaussian state restriction might be analytically and numerically tractable \cite{PRXQuantum.5.010313}.

Solving for most likely paths or optimal control can be hard, even numerically.  The costate operator leads to a twofold increase in the problem's dimensionality.  The model reduction strategy we have presented is helpful for a monitored oscillator with a quadratic control Hamiltonian. In this case, instead of the costate operator,  we track  10 additional scalars that depend on the commutator and anti-commutator of the state and the costate.  It will be worthwhile to explore other model reduction strategies for finding optimal control. On this note, neural network-based methods have been adopted to optimize state preparation schemes \cite{PRXQuantum.6.010321} and feedback control protocols \cite{PRXQuantum.4.030305,reuer_realizing_2023, vaidhyanathan2024quantumfeedbackcontroltransformer,PhysRevLett.134.020601}. 

Finally, the conditional stochastic evolution considered here is Markovian. Markovian approximation is valid when detector correlation timescales are much shorter than the timescale associated with the system's dynamics. In such a situation, the memory effects of the system-detector interaction are negligible.  It will be insightful to explore the CDJ-Pontryagin formalism for non-Markovian dynamics. A pseudo-Lindbladian master equation-based description can be helpful in this scenario \cite{ Koch_2016,RevModPhys.89.015001, Groszkowski2023simplemaster}. In conclusion, our work provides a systematic strategy to find optimal control for systems undergoing general nonlinear dynamics and offers several possible avenues for further exploration.

\section*{\label{sec:ack}Acknowledgement}
We thank Philippe Lewalle, Adithi Ajith, K.~Birgitta Whaley, Alain Sarlette, Yipei Zhang, Sreenath Manikandan, Ryotatsu Yanagimoto,   Bibek Bhandari, Alok Nath Singh, Tryphon T.~Georgiou, and Olga Movilla for helpful discussions. We acknowledge support from the John Templeton Foundation Grant ID 63209 and the Army Research Office grant W911NF-22-1-0258.

\printbibliography
  
\appendix
\section{\label{appA0_strat}Variation of the stochastic action}
In this Appendix, we look at the variational extremization of the action given by Eqs.~\eqref{stoch_h_general} and \eqref{stoch_S_general}.  We consider trajectories with initial \begin{widetext} state  $\hat{\rho}_i$ at $t=0$ and final state $\hat{\rho}_f$ at time $t=t_f$. The variation of the stochastic action due to variations in $\hat{\rho}$, $\hat{\sigma}$, and $r$ is
\begin{equation}
\begin{split}
    \delta \mathcal{S}&=\int_0^{t_f}dt\left(-\textrm{Tr}\left(\delta\hat{\sigma}\frac{\partial \hat{\rho}}{\partial t}+\hat{\sigma}\frac{\partial \delta\hat{\rho}}{\partial t}\right)+\delta \mathcal{H}\right)=\int_0^{t_f}dt(\delta \Psi+\delta \mathcal{H}),
    \end{split}
\end{equation}
where $\delta \Psi = -\textrm{Tr}\left(\delta\hat{\sigma}\frac{\partial \hat{\rho}}{\partial t}+\hat{\sigma}\frac{\partial \delta\hat{\rho}}{\partial t}\right)$.
Since we look at trajectories with fixed initial and final states,  $\delta \hat{\rho}=0$ at the boundaries. Thus, we can write
\begin{equation}
   \int_0^{t_f}dt \delta \Psi = \int_0^{t_f}dt \textrm{Tr}\left(\frac{\partial\hat{\sigma}}{\partial t}\delta\hat{\rho}-\delta\hat{\sigma}\frac{\partial \hat{\rho}}{\partial t}\right).
   \label{del_psi}
\end{equation}
Variation of the stochastic Hamiltonian is given by
\begin{equation}
\begin{split}
\delta\mathcal{H}&=\delta\textrm{Tr}\left(\hat{\sigma}\mathfrak{F}(\hat{\rho},r)\right)-\frac{1}{2\tau}\delta\left(r^2-2r\expval{\hat{L}}+\expval{\hat{L}^2}\right).
\end{split}
\end{equation}

We can derive the following identities

\begin{subequations}
\begin{align}
\begin{split}
\delta\textrm{Tr}\left(\hat{\sigma}\left(-i[\hat{H},\hat{\rho}]\right)\right)&=\textrm{Tr}\Big(i[\hat{H},\hat{\sigma}]\delta\hat{\rho}-i\delta\hat{\sigma}[\hat{H},\hat{\rho}]\Big),
\end{split}\\
\begin{split}
\delta\textrm{Tr}\left(-\hat{\sigma}\frac{1}{4\tau}\left[\hat{L}^2-\expval{\hat{L}^2},\hat{\rho}\right]_+\right)  & = \textrm{Tr}\Big(-\delta\hat{\sigma} \frac{1}{4\tau}\left[\hat{L}^2-\expval{\hat{L}^2},\hat{\rho}\right]_+\\&- \frac{1}{4\tau}\left(\left[\hat{L}^2-\expval{\hat{L}^2},\hat{\sigma}\right]_+-2\expval{\hat{\sigma}}\hat{L}^2\right)\delta\hat{\rho}\Big),
\end{split}\\
\begin{split}
    \delta \textrm{Tr}\Bigg(\hat{\sigma}\frac{1}{2\tau}r\Big[\hat{L}-\expval{\hat{L}},\hat{\rho}\Big]_+\Bigg) &= 
    \frac{\delta r}{2\tau}\expval{\Big[\hat{L}-\expval{\hat{L}},\hat{\sigma}\Big]_+}\\&+ \textrm{Tr}\Big(\frac{\delta \hat{\sigma}}{2\tau}r\Big[\hat{L}-\expval{\hat{L}},\hat{\rho}\Big]_++\frac{r}{2\tau}\left(\Big[\hat{L}-\expval{\hat{L}},\hat{\sigma}\Big]_+-2\expval{\hat{\sigma}}\hat{L}\right)\delta\hat{\rho}\Big)
\end{split}\\
\begin{split}
    -\frac{1}{2\tau}\delta\left(r^2-2r\expval{\hat{L}}+\expval{\hat{L}^2}\right)&=-\frac{\delta r}{\tau}(r-\expval{\hat{L}})-\frac{1}{2\tau}\textrm{Tr}\left((-2r\hat{L}+\hat{L}^2)\delta\hat{\rho}\right).
\end{split}
\end{align}
\label{del_H}
\end{subequations}

Using Eqs.~\eqref{del_psi} and \eqref{del_H}, we can write the variation of the stochastic action in the following form
\begin{equation}
   \delta \mathcal{S}=\int_0^{t_f}dt\Bigg(\frac{\delta r}{\tau}\left(\frac{1}{2}\expval{\Big[\hat{L}-\expval{\hat{L}},\hat{\sigma}\Big]_+}-\left(r-\expval{\hat{L}}\right)\right)+\textrm{Tr}\left(\delta\hat{\sigma}(\hat{\Gamma}_1-\frac{\partial \hat{\rho}}{\partial t})+(-\hat{\Gamma}_2+\frac{\partial \hat{\sigma}}{\partial t})\delta\hat{\rho}\right)\Bigg),
\end{equation}

where 
\begin{subequations}
\begin{align}
    \begin{split}
        \hat{\Gamma}_1 &= -i[\hat{H},\hat{\rho}] -\frac{1}{4\tau}\Big[\Delta\hat{V},\hat{\rho}\Big]_++\frac{r}{2\tau}\Big[\Delta\hat{L},\hat{\rho}\Big]_+,
    \end{split}\\
    \begin{split}
        \hat{\Gamma}_2 &= -i[\hat{H},\hat{\sigma}] +\frac{1}{4\tau}\left(\left[\Delta\hat{V},\hat{\sigma}\right]_+-2\expval{\hat{\sigma}}\hat{V}\right)-\frac{r}{2\tau}\left(\Big[\Delta\hat{L},\hat{\sigma}\Big]_+-2\expval{\hat{\sigma}}\hat{L}\right)+\frac{1}{2\tau}(-2r\hat{L}+\hat{L}^2).
    \end{split}
\end{align}
\label{gamma12}
\end{subequations}For most likely paths, $\delta \mathcal{S}=0$ for arbitrary variations of the state, costate, and readout. Therefore,  the optimal readout is given by
\begin{equation}
    r^\star =\expval{\hat{L}}+u=\expval{\hat{L}}+ \frac{1}{2}\expval{\Big[\Delta\hat{L},\hat{\sigma}\Big]_+},
\end{equation}
where $u$ is the optimal noise. Substituting the value of the optimal readout in Eq.~\eqref{gamma12}, we get 
\begin{subequations}
\begin{align}
    \begin{split} \hat{\Gamma}_1^{\textrm{opt}} &=-i[\hat{H},\hat{\rho}] -\frac{1}{4\tau}\Big[\Delta\hat{V},\hat{\rho}\Big]_++\frac{r^\star}{2\tau}\Big[\Delta\hat{L},\hat{\rho}\Big]_+,
    \end{split}\\
    \begin{split}
\hat{\Gamma}_2^{\textrm{opt}} &= -i[\hat{H},\hat{\sigma}] +\frac{1}{4\tau}\left(\left[\Delta\hat{V},\hat{\sigma}\right]_+-2(\expval{\hat{\sigma}}-1)\hat{V}\right)-\frac{r^\star}{2\tau}\left(\Big[\Delta\hat{L},\hat{\sigma}\Big]_+-2(\expval{\hat{\sigma}}-1)\hat{L}\right).
    \end{split}
\end{align}
\label{gamma12_opt}
\end{subequations}The equations for the optimal paths are 
\begin{equation}
    \frac{\partial\hat{\rho}}{\partial t}=\hat{\Gamma}_1^{\textrm{opt}},\quad \frac{\partial\hat{\sigma}}{\partial t}=\hat{\Gamma}_2^{\textrm{opt}}
    \label{OPeq_appnx}
\end{equation}The optimal Hamiltonian $\mathcal{H}^\star(\hat{\sigma},\hat{\rho})=\mathcal{H}(\hat{\sigma},\hat{\rho}, r^\star)$ is given by
\begin{equation}
    \begin{split}
\mathcal{H}^\star(\hat{\sigma},\hat{\rho}) = \expval{i[\hat{H},\hat{\sigma}]-\frac{1}{4\tau}\left[\Delta\hat{V},\hat{\sigma}\right]_+}+\frac{r^\star}{\tau}u-\frac{1}{2\tau}\left(u^2+\textrm{Var}(\hat{L})\right).
    \end{split}
\end{equation}
\end{widetext}
We can always redefine $\hat{\sigma}\equiv\hat{\sigma}+(1-\expval{\hat{\sigma}})\hat{\mathds{1}}$ such that $\expval{\hat{\sigma}}=1$ at all times. Then, the optimal readout is given by
\begin{equation}
    r^\star =\frac{1}{2}\expval{\Big[\hat{L},\hat{\sigma}\Big]_+}.\label{opt_readout_appendix}
\end{equation}
%\end{widetext}
The equations for optimal paths take the form
\begin{equation}
    \begin{split}
        \frac{\partial\hat{\rho}}{\partial t}&=-i[\hat{H},\hat{\rho}] -\frac{1}{4\tau}\Big[\Delta\hat{V},\hat{\rho}\Big]_++\frac{r^\star}{2\tau}\Big[\Delta\hat{L},\hat{\rho}\Big]_+,\\\frac{\partial\hat{\sigma}}{\partial t}&= -i[\hat{H},\hat{\sigma}] +\frac{1}{4\tau}\left[\Delta\hat{V},\hat{\sigma}\right]_+-\frac{r^\star}{2\tau}\Big[\Delta\hat{L},\hat{\sigma}\Big]_+.
    \end{split}
    \label{op_reduced}
\end{equation}
The optimal Hamiltonian is 
\begin{equation}
    \begin{split}
\mathcal{H}^\star(\hat{\sigma},\hat{\rho}) &=\expval{i[\hat{H},\hat{\sigma}]}+\frac{1}{8\tau}\expval{\left[\Delta\hat{L},\hat{\sigma}\right]_+}^2\\&-\frac{1}{4\tau}\expval{\left[(\Delta\hat{L})^2,\hat{\sigma}\right]_+}\\&= \expval{i[\hat{H},\hat{\sigma}]}+\frac{1}{8\tau}\expval{\left[\hat{L},\hat{\sigma}\right]_+}^2\\&-\frac{1}{4\tau}\expval{\left[\hat{V},\hat{\sigma}\right]_+}.
    \end{split}
\end{equation}
The optimal path equations in Eq.~\eqref{op_reduced} can be found from the optimal Hamiltonian and can be expressed as 
\begin{equation}
    \frac{\partial\hat{\rho}}{\partial t}=\frac{\delta\mathcal{H}^\star}{\delta\hat{\sigma}},\quad \frac{\partial\hat{\sigma}}{\partial t}=-\frac{\delta\mathcal{H}^\star}{\delta\hat{\rho}},
\end{equation}
where the derivatives above denote functional derivatives.
\section{\label{CDJ_vs_costate}Relationship between the costate operator and CDJ auxiliary variables}
In this Appendix, we clarify the relationship between the costate operator and CDJ auxiliary variables (see Ref.~\cite{zhang2025optimalschedulemultichannelquantum} for further discussions). For simplicity, we look into the example of a single monitored qubit. We will see that the costate operator provides constraints on the qubit's degrees of freedom. Suppose we express the state of the qubit in terms of a column vector of  Bloch coordinates $\pmb{q}=\left(q_1, q_2, q_3\right)^\top$. In other words, the qubit density matrix is given by
\begin{equation}
    \hat{\rho}=\frac{1}{2}\left(\hat{\mathds{1}}+q_1\hat{\sigma}_x+q_2\hat{\sigma}_y+q_3\hat{\sigma}_z\right),
\end{equation}
where $\hat{\sigma}_{x,y, z}$ above denote the Pauli operators. Suppose the qubit dynamics is given by a general stochastic master equation (like Eq.~\eqref{rho_evol_append})
\begin{equation}
    \frac{\partial \hat{\rho}}{\partial t}=\hat{\mathfrak{F}}_{\hat{\rho}}[\hat{\rho},r],
    \label{qubit_SME}
\end{equation} where $r$ denotes the measurement readouts. Eq.~\eqref{qubit_SME} can be expressed in terms of Bloch coordinates by taking the trace with respect to the Pauli operators. In other words, \begin{equation}
\dot{q}_1=\textrm{Tr}\left(\hat{\sigma}_x\hat{\mathfrak{F}}_{\hat{\rho}}[\hat{\rho},r]\right)=\mathcal{F}_1(\pmb{q},r),
\label{SME_bloch_defn}
\end{equation} and so on. Then, the qubit dynamics in terms of Bloch coordinates can be expressed as
\begin{equation}
    \dot{\pmb{q}}=\pmb{\mathcal{F}}(\pmb{q},r),
\end{equation}
If the conditional probability of readout $r$ in time $dt$ is given by $P(r|\pmb{q})\sim e^{dt\mathcal{G}(\pmb{q}, r)}$ (see Sec.~\ref{stoch_H_CDJ}), the CDJ stochastic Hamiltonian can be written as \cite{PhysRevA.92.032125}
\begin{equation}
    \mathcal{H}(\pmb{p}, \pmb{q}, r)=\pmb{p}^{\top}\cdot\pmb{\mathcal{F}}(\pmb{q}, r)+\mathcal{G}(\pmb{q}, r),
    \label{cdj_h_cdj_aux}
\end{equation}
where the column vector $\pmb{p}=\left(p_1, p_2, p_3\right)^\top$ denotes the CDJ auxiliary variables corresponding to the Bloch coordinates. Like in Sec.~\ref{stoch_H_CDJ},  $\mathcal{G}(\pmb{q},r) \equiv \textrm{ln}\,P(r|\pmb{q})$ denotes the log probability of  readouts. From the Hamiltonian in Eq.~\eqref{cdj_h_cdj_aux}, the most likely dynamics can be found using Eq.~\eqref{OP_eq_parametrized}.

Next, we examine the (Hermitian) costate operator $\hat{\sigma}$  based description. Since any qubit operator can be expressed in terms of Paulis, we can write
\begin{equation}
\hat{\sigma}=\lambda_0\hat{\mathds{1}}+\lambda_1\hat{\sigma}_x+\lambda_2\hat{\sigma}_y+\lambda_3\hat{\sigma_z}.\label{eq_sigma_expand}
\end{equation}
When the log probability is the cost function, the general Pontryagin Hamiltonian is given by (see Eq.~\eqref{H_pontrya_ctsms})
\begin{equation}
    \mathcal{H}(\hat{\sigma},\hat{\rho}, r)=\textrm{Tr}\left(\hat{\sigma}\hat{\mathfrak{F}}_{\hat{\rho}}[\hat{\rho},r]\right)+\mathcal{G}(\hat{\rho},r),
\end{equation}
where $\mathcal{G}(\hat{\rho},r)$ expresses the log probability of readouts in terms of the density matrix. Now using the expansion in Eq.~\eqref{eq_sigma_expand}, the definition in Eq.~\eqref{SME_bloch_defn} and the fact that  $\textrm{Tr}\left(\hat{\mathfrak{F}}_{\hat{\rho}}[\hat{\rho},r]\right)=0$ since Eq.~\eqref{qubit_SME} is trace preserving, we can write
\begin{equation}
    \mathcal{H}(\hat{\sigma},\hat{\rho}, r)=\sum_i\lambda_i\mathcal{F}_i(\pmb{q},r)+\mathcal{G}(\pmb{q},r).
\end{equation}
 The above expression has the same form as Eq.~\eqref{cdj_h_cdj_aux}. Thus, the CDJ auxiliary variables correspond to a parametrization of the costate operator. Furthermore, the costate operator constraints different degrees of motion of the system.
\section{\label{appA2_strat}Time evolution identities}
This Appendix provides formulae for the time evolution of different expectation values along the optimal paths. Assume $\hat{B}_\chi$ is an explicitly time-dependent observable, in general dependent on control  $\chi$ as well. Similarly, we introduce the subscript $\chi$ to denote the control dependence of the observables and the optimal readouts in  Eqs.~\eqref{opt_readout_appendix} and \eqref{op_reduced}.  We also define the anti-commutator and the commutator of the state and costate as 
\begin{equation}
\hat{\Omega}=\frac{1}{2}\left[\hat{\rho},\hat{\sigma}\right]_+,\quad \hat{\Lambda}= \left[\hat{\rho},\hat{\sigma}\right].
    \label{omega}
\end{equation}
Note that $\textrm{Tr}\:\hat{\Omega}= 1$ and $\textrm{Tr}\:\hat{\Lambda}=0$. We can show the following identities
\begin{widetext}
\begin{subequations}
\begin{align}
    \begin{split}
      \label{eq_b_derivative}\frac{d\expval{\hat{B}_\chi}}{dt}&=\expval{\frac{\partial\hat{B}_\chi}{\partial t}+\dot{\chi}\frac{\partial\hat{B}_\chi}{\partial \chi}} +\Bigg\langle i\left[\hat{H}_\chi,\hat{B}_\chi\right]-\frac{1}{4\tau}\left[\Delta\hat{V}_\chi,\hat{B}_\chi\right]_++\frac{r_\chi}{2\tau}\Big[\Delta\hat{L}_\chi,\hat{B}_\chi\Big]_+\Bigg\rangle\\&=\expval{\frac{\partial\hat{B}_\chi}{\partial t}+\dot{\chi}\frac{\partial\hat{B}_\chi}{\partial \chi}+i\left[\hat{H}_\chi,\hat{B}_\chi\right] +\frac{1}{2\tau}\expval{\Delta\hat{L}_\chi^2}\hat{B}_\chi}\\&+\Bigg\langle -\frac{1}{4\tau}\left[\Delta\hat{L}_\chi^2,\hat{B}_\chi\right]_++\frac{r_\chi-\expval{\hat{L}_\chi}}{2\tau}\Big[\Delta\hat{L}_\chi,\hat{B}_\chi\Big]_+\Bigg\rangle\\&=\expval{\frac{\partial\hat{B}_\chi}{\partial t}+\dot{\chi}\frac{\partial\hat{B}_\chi}{\partial \chi}+i\left[\hat{H}_\chi,\hat{B}_\chi\right] }+\Bigg\langle -\frac{1}{4\tau}\left[\Delta\hat{L}_\chi^2,\Delta\hat{B}_\chi\right]_++\frac{r_\chi-\expval{\hat{L}_\chi}}{2\tau}\Big[\Delta\hat{L}_\chi,\Delta\hat{B}_\chi\Big]_+\Bigg\rangle,
    \end{split}\\
    \begin{split}
      \frac{d\textrm{Tr}(\hat{B}_\chi\hat{\Omega})}{dt}&=\textrm{Tr}\left(\left(\frac{\partial\hat{B}_\chi}{\partial t}+\dot{\chi}\frac{\partial\hat{B}_\chi}{\partial \chi}+i\left[\hat{H}_\chi,\hat{B}_\chi\right]\right)\hat{\Omega} \right)+\frac{1}{8\tau}\textrm{Tr}\left(\left[\hat{V}_\chi,\hat{B}_\chi\right]\hat{\Lambda}\right)-\frac{r_\chi}{4\tau}\textrm{Tr}\left(\left[\hat{L}_\chi,\hat{B}_\chi\right]\hat{\Lambda}\right),
      \label{b_Omega_strat}
    \end{split}\\
    \begin{split}
      \frac{d\textrm{Tr}(\hat{B}_\chi\hat{\Lambda})}{dt}&=\textrm{Tr}\left(\left(\frac{\partial\hat{B}_\chi}{\partial t}+\dot{\chi}\frac{\partial\hat{B}_\chi}{\partial \chi}+i\left[\hat{H}_\chi,\hat{B}_\chi\right]\right)\hat{\Lambda} \right) +\frac{1}{2\tau}\textrm{Tr}\left(\left[\hat{V}_\chi,\hat{B}_\chi\right]\hat{\Omega}\right)-\frac{r_\chi}{\tau}\textrm{Tr}\left(\Big[\hat{L}_\chi,\hat{B}_\chi\Big]\hat{\Omega}\right),
\label{b_Lambda_strat} \end{split}
    \end{align}
\label{omega_lambda_evol_eqns}
\end{subequations}where $\Delta \hat{O}=\hat{O}-\expval{\hat{O}}$, for any observable $\hat{O}$.   Note that $\textrm{Tr}(\hat{L}_\chi\hat{\Omega})=r_\chi$. If $\hat{B}_\chi=\hat{L}_\chi$ with no explicit time dependence, Eq.~\eqref{b_Omega_strat} leads to 
\begin{equation}
    \begin{split}
      \frac{dr_\chi}{dt}&=\textrm{Tr}\left(\left(\dot{\chi}\frac{\partial\hat{L}_\chi}{\partial \chi}+i\left[\hat{H}_\chi,\hat{L}_\chi\right]\right)\hat{\Omega} \right).   
      \label{L_omega0}
    \end{split}
\end{equation}
It follows that under quantum nondemolition measurements, with no control, $\frac{dr}{dt}=0$. Thus, the most likely readout is constant for quantum nondemolition measurements.

For arbitrary $\hat{A}_\chi$ $\hat{B}_\chi$  operators, we adopt the following notations
\begin{equation}
    \hat{\mathcal{C}}^{AB}_\chi=\frac{1}{2}\left[\Delta\hat{A}_\chi,\Delta\hat{B}_\chi\right]_+, \quad \textrm{therefore, }  \quad \textrm{Cov}(\hat{A}_\chi,\hat{B}_\chi)=\expval{\hat{\mathcal{C}}^{AB}_\chi}.
\end{equation}
We also denote
\begin{equation}
    \frac{\partial}{\partial t}+\dot{\chi}\frac{\partial}{\partial \chi}=\partial_\eta,\quad \hat{N}_\chi=-\frac{1}{4\tau}\Delta\hat{L}_\chi^2+\frac{r_\chi-\hat{L}_\chi}{2\tau}\Delta\hat{L}_\chi.
\end{equation}
We can show that
    \begin{equation}
\begin{split}
    \frac{d \expval{\hat{\mathcal{C}}^{AB}_\chi}}{dt}&=\expval{\partial_\eta\hat{\mathcal{C}}^{AB}_\chi+i[\hat{H}_\chi,\hat{\mathcal{C}}^{AB}_\chi]+\frac{\expval{\Delta\hat{L}_\chi^2}}{2\tau}\hat{\mathcal{C}}^{AB}_\chi}+\expval{\left[\hat{N}_\chi,\hat{\mathcal{C}}^{AB}_\chi\right]_+}\\&=\textrm{Cov}(\partial_\eta\hat{A}_\chi,\hat{B}_\chi)+\textrm{Cov}(\hat{A}_\chi,\partial_\eta\hat{B}_\chi)+\expval{i[\hat{H}_\chi,\hat{\mathcal{C}}^{AB}_\chi]+\frac{\expval{\Delta\hat{L}_\chi^2}}{2\tau}\hat{\mathcal{C}}^{AB}_\chi + \left[\hat{N}_\chi,\hat{\mathcal{C}}^{AB}_\chi\right]_+}.
\end{split}
\end{equation} 
The above equation is consistent with $\expval{\hat{B}_\chi}$ evolution in Eq.~\eqref{eq_b_derivative}.

\section{\label{app:cdjp_SHO}Optimal path for the quantum harmonic oscillator}For a simple harmonic oscillator with the Hamiltonian given by  Eq.~\eqref{hamiltonian_sho}, and quadratures Eqs.~\eqref{l_theta} and \eqref{m_theta}, we can write
\begin{equation}
    \hat{H}=\frac{1}{2}\left(\hat{L}_\theta^2+\hat{M}_\theta^2\right).
\end{equation}
The following identities are true
\begin{equation}
    \begin{split}
    &\left[\hat{L}_\theta,\hat{M}_\theta\right]=i,\quad \left[\hat{H},\hat{L}_\theta\right]=-i\hat{M}_\theta,\quad \left[\hat{H},\hat{M}_\theta\right]=i\hat{L}_\theta,\\
&\left[\hat{L}_\theta^2,\hat{M}_\theta\right]=2i\hat{L}_\theta,\quad \frac{\partial \hat{L}_\theta}{\partial \theta}=\hat{M}_\theta, \quad \frac{\partial \hat{M}_\theta}{\partial \theta}=-\hat{L}_\theta.
    \end{split}
\end{equation}
From Eq.~\eqref{opt_readout_appendix}  and \eqref{omega}, the optimal readout for adaptive quadrature measurement can be written as
\begin{equation}
    r_\theta=\frac{1}{2}\expval{\Big[\hat{L}_\theta,\hat{\sigma}\Big]_+}=\textrm{Tr}(\hat{L}_\theta\hat{\Omega}).
\end{equation}
From Eq.~\eqref{L_omega0}, we can derive
\begin{equation}
   \frac{dr_\theta}{dt}=\dot{\phi}\textrm{Tr}\left(\hat{M}_\theta\hat{\Omega} \right),
\end{equation}
where $\dot{\phi}=\dot{\theta}+1$. Now we assume $v_\theta = \textrm{Tr}\left(\hat{M}_\theta\hat{\Omega} \right)$. The evolution of $v_\theta$ can be derived from Eq.~\eqref{b_Omega_strat} to be
\begin{equation}
\begin{split}
    \frac{dv_\theta}{dt}&=-\dot{\phi}r_\theta+\frac{i}{4\tau}\textrm{Tr}\left(\hat{L}_\theta\hat{\Lambda} \right).
\end{split}
\end{equation}
We denote $w_\theta=i\expval{\left[\hat{\sigma},\hat{L}_\theta\right]}=i\textrm{Tr}\left(\hat{L}_\theta\hat{\Lambda}\right)$. From Eq.~\eqref{b_Lambda_strat}, we get
\begin{equation}
    \frac{dw_\theta}{dt}=i\dot{\phi}\textrm{Tr}\left(\hat{M}_\theta\hat{\Lambda}\right).
\end{equation}
Similarly, with the definition $z_\theta=i\expval{\left[\hat{\sigma},\hat{M}_\theta\right]}=i\textrm{Tr}\left(\hat{M}_\theta\hat{\Lambda}\right)$, we get
\begin{equation}
    \frac{dz_\theta}{dt}=-\dot{\phi}w_\theta.
\end{equation}
Thus, the evolution of the optimal noise can be expressed in terms of the following coupled equations
\begin{equation}
    \frac{dr_\theta}{dt}=\dot{\phi}v_\theta, \quad \frac{dv_\theta}{dt}=-\dot{\phi}r_\theta+\frac{w_\theta}{4\tau},\quad \frac{dw_\theta}{dt}=\dot{\phi}z_\theta,\quad \frac{dz_\theta}{dt}=-\dot{\phi}w_\theta,
    \label{uvwz_eq_app}
\end{equation}
with $\dot{\phi}=\dot{\theta}+1$.
\section{\label{numerics_theta0}Numerical Methods for monitored oscillator undergoing position measurements}Here, we consider the numerical methods for finding the optimal paths for oscillators undergoing position measurements (i.e.~$\theta(t)=0$ in \eqref{l_theta}, see Figure~\ref{tmpfig}). The most likely state evolution in Eq.~\eqref{op_rho_sho} can be expressed as 
\begin{equation}
    \frac{\partial \hat{\rho}}{\partial t}=\hat{F}\hat{\rho}+\hat{\rho}\hat{F}^\dagger,
    \label{rho_eq_theta0_compact}
\end{equation}
with $\hat{F}=-i\hat{H}-\frac{1}{4\tau}\Delta\hat{V}_\theta+\frac{r_\theta}{2\tau}\Delta\hat{L}_\theta$. We take $r_\theta$ to be of the form given by Eq.~\eqref{r_th_expr}, with the complex constant $\alpha$ and real constants $A$ and $B$ to be determined. To integrate Eq.~\eqref{rho_eq_theta0_compact}, $\hat{\rho}(t+dt)$ is expressed as 
\begin{equation}
    \hat{\rho}(t+dt)=\frac{\hat{\rho}^\prime}{\textrm{Tr}\hat{\rho}^\prime},
\end{equation}
with 
\begin{equation}
\hat{\rho}^\prime=\hat{\mathcal{U}}_{dt}\hat{M}_{dt}\hat{\rho}\hat{M}_{dt}^\dagger\hat{\mathcal{U}}^\dagger_{dt},
\end{equation}
where $\hat{M}_{dt}$ is the Stratonovich form of the Kraus operator
\begin{equation}
    \hat{M}_{dt}=\hat{\mathds{1}}+\frac{rdt}{2\tau}\hat{L}-\frac{dt}{4\tau}\hat{L}^2.
\end{equation}
The above form ensures the positivity of the density matrix throughout the simulation \cite{appelo2024krauskinghighordercompletely}. We find $\alpha$, $A$ and $B$ (see Eq.~\eqref{r_th_expr}) by maximizing the fidelity between the target final state $\hat{\rho}_f$ and the simulated final state $\hat{\rho}(t_f)$, defined as
\begin{equation}
\mathcal{F}(t_f, \hat{\rho}_f)=\textrm{Tr}\left(\hat{\rho}(t_f)\hat{\rho}_f\right).
\label{def_fidelity}
\end{equation}
We can adopt the above definition of fidelity since we restrict our analysis to pure states. 
\section{\label{app_parametric}Relevant equations for a monitored parametric oscillator}
We recall McCoy's formula for position and momentum commutation \cite{McCoy1929OnCF}
\begin{equation}
    \left[\hat{X}^n,\hat{P}^m\right]=\sum_{k=1}C(n,m,k)\hat{X}^{n-k}\hat{P}^{m-k},
    \label{xnpm_commute}
\end{equation}
where 
\begin{equation}
    C(n,m,k) =\begin{cases}
        \frac{-(-i)^kn!m!}{k!(n-k)!(m-k)!}\quad &\textrm{for }k\leq n,m\\ 
         0 \quad &\textrm{otherwise}.
    \end{cases}
\end{equation}
From now on, we will assume the sum in Eq.~\eqref{xnpm_commute} starts from $k=1$ without explicitly mentioning it. For a Hamiltonian
\begin{equation}
    \hat{H}=\frac{1}{2}\left(\hat{X}^2+\hat{P}^2\right)+\lambda_1(t)\hat{X}^2+\lambda_2(t)\hat{X}^3,
\end{equation}
and $\hat{L}_\theta$ defined 
in Eq.~\eqref{l_theta}, the following identities are true.

\begin{subequations}
\begin{align}
    \begin{split}
\left[\hat{H},\hat{X}^n\hat{P}^m\right]&=\sum_k\Bigg(\frac{1}{2}(1+2\lambda_1)C(2,m,k)\hat{X}^{n+2-k}\hat{P}^{m-k}+\lambda_2C(3,m,k)\hat{X}^{n+3-k}\hat{P}^{m-k}\\&-\frac{1}{2}C(n,2,k)\hat{X}^{n-k}\hat{P}^{m+2-k},\Bigg),
    \label{Hxnpm_commute}
\end{split}\\
\begin{split}
\left[\hat{L}_\theta,\hat{X}^n\hat{P}^m\right]&=\sum_{k}\Bigg(
\cos\theta C(1,m,k)\hat{X}^{n+1-k}\hat{P}^{m-k}-\sin\theta C(n,1,k)\hat{X}^{n-k}\hat{P}^{m+1-k}\Bigg),
    \label{Lxnpm_commute}
\end{split}\\
\begin{split}
\left[\hat{L}^2_\theta,\hat{X}^n\hat{P}^m\right]&=\sum_k\Bigg(\cos^2\theta C(2,m,k)\hat{X}^{n+2-k}\hat{P}^{m-k}\\&+2\sin\theta\cos\theta\Big(
 C(1,m,k)- C(n,1,k)\Big)\hat{X}^{n+1-k}\hat{P}^{m+1-k}-\sin^2\theta C(n,2,k)\hat{X}^{n-k}\hat{P}^{m+2-k},\Bigg).
    \label{L2xnpm_commute}
\end{split}
\end{align}
\label{XnPm_commute_eqns}
\end{subequations}
We define the following scalars 
\begin{equation}
   % \begin{split}
\Gamma(n,m)=\textrm{Tr}\left(\hat{X}^n\hat{P}^m\hat{\Omega}\right),\quad \kappa(n,m)=i\textrm{Tr}\left(\hat{X}^n\hat{P}^m\hat{\Lambda}\right).
\end{equation}
Using Eqs.~\eqref{omega_lambda_evol_eqns} and \eqref{XnPm_commute_eqns}, we can derive the following recurrence relations
\begin{equation}
    \begin{split}
        \frac{d\Gamma(n,m)}{dt}&=i\sum_{k=1}\Bigg[\Bigg(\frac{1}{2}(1+2\lambda_1)C(2,m,k)\Gamma(n+2-k,m-k)+\lambda_2C(3,m,k)\Gamma(n+3-k,m-k)\\&-\frac{1}{2}C(n,2,k)\Gamma(n-k,m+2-k),\Bigg)-\frac{1}{8\tau}\Bigg(\cos^2\theta C(2,m,k)\kappa(n+2-k,m-k)\\&+2\sin\theta\cos\theta\Big(
 C(1,m,k)- C(n,1,k)\Big)\kappa(n+1-k,m+1-k)\\&-\sin^2\theta C(n,2,k)\kappa(n-k,m+2-k),\Bigg)+\frac{r_\theta}{4\tau}\Bigg(
\cos\theta C(1,m,k)\kappa(n+1-k,m-k)\\&-\sin\theta C(n,1,k)\kappa(n-k,m+1-k)\Bigg)\Bigg],
 \label{dGamnmdt}
    \end{split}
\end{equation}
\begin{equation}
    \begin{split}
   \frac{d\kappa(n,m)}{dt}&=i\sum_{k=1}\Bigg[\Bigg(\frac{1}{2}(1+2\lambda_1)C(2,m,k)\kappa(n+2-k,m-k)+\lambda_2C(3,m,k)\kappa(n+3-k,m-k)\\&-\frac{1}{2}C(n,2,k)\kappa(n-k,m+2-k),\Bigg)+\frac{1}{2\tau}\Bigg(\cos^2\theta C(2,m,k)\Gamma(n+2-k,m-k)\\&+2\sin\theta\cos\theta\Big(
 C(1,m,k)- C(n,1,k)\Big)\Gamma(n+1-k,m+1-k)\\&-\sin^2\theta C(n,2,k)\Gamma(n-k,m+2-k),\Bigg)-\frac{r_\theta}{\tau}\Bigg(
\cos\theta C(1,m,k)\Gamma(n+1-k,m-k)\\&-\sin\theta C(n,1,k)\Gamma(n-k,m+1-k)\Bigg)\Bigg],
 \label{dkapnmdt}
    \end{split}
\end{equation}

with $\Gamma(0,0)=1$ and $\kappa(0,0)=0$. Also, the optimal readout $r_\theta$ can be written as
\begin{equation}
    r_\theta=\cos\theta\Gamma(1,0)+\sin\theta\Gamma(0,1).
\end{equation}
We define the order of $\Gamma(n,m)$ or $\kappa(n,m)$ to be $n+m$. Note that in Eqs.~\eqref{dGamnmdt} \eqref{dkapnmdt}, only the terms multiplied by $\lambda_2$ lead to an increase in the order (since $k\geq 1$). Therefore, the model reduction strategy presented in Secs.~\ref{sho_op_strato} and \ref{control_SHO} only works when the oscillator does not have any anharmonic potential.  We can write the optimal Hamiltonian from Eq.~\eqref{h_op_general} as
\begin{equation}
    \begin{split}
\mathcal{H}^\star(\hat{\sigma},\hat{\rho},\theta,\lambda_1,\lambda_2) &= -i\textrm{Tr}\left(\hat{H}\hat{\Lambda}\right)+\frac{1}{2\tau}\left(\textrm{Tr}\left(\hat{L}_\theta\hat{\Omega}\right)\right)^2-\frac{1}{2\tau}\textrm{Tr}\left(\hat{L}_\theta^2\hat{\Omega}\right),
    \end{split}
\end{equation}
which can be expanded as 
\begin{equation}
\begin{split}
        \mathcal{H}^\star&=-\Big(\frac{1}{2}(1+2\lambda_1)\kappa(2,0)+\frac{1}{2}\kappa(0,2)+\lambda_2\kappa(3,0)\Big)\\&+\frac{1}{2\tau}\Big(\cos\theta\Gamma(1,0)+\sin\theta\Gamma(0,1)\Big)^2-\frac{1}{2\tau}\Big(\cos^2\theta\Gamma(2,0)+\cos\theta\sin\theta(2\Gamma(1,1)-i)+\sin^2\theta\Gamma(0,2)\Big).
        \label{hstar_1}
\end{split}
\end{equation}
Assuming $\lambda_1\in[-\lambda_{1}^{\max},\lambda_{1}^{\max}]$ and $\lambda_2\in[-\lambda_{2}^{\max},\lambda_{2}^{\max}]$, according to Condition I of the PMP presented in  Eq.~\eqref{PMP_cond_hmax}, 
the optimal values of these parameters are 
\begin{equation}
\begin{split}
    \lambda_1=-\lambda_1^{\max}\textrm{sign}(\kappa(2,0)),\quad  \lambda_2 =-\lambda_2^{\max}\textrm{sign}(\kappa(3,0)).
\end{split}
\label{l1l2_optimal_cond}
\end{equation} Thus, we get a bang-bang form of optimal control. Eq.~\eqref{hstar_1} then becomes
\begin{equation}
   \begin{split}
        \mathcal{H}^\star&=\lambda_1^{\max}|\kappa(2,0)|+\lambda_2^{\max}|\kappa(3,0)|-\frac{1}{2}\Big(\kappa(2,0)+\kappa(0,2)\Big)\\&+\frac{1}{4\tau}\Big(\Gamma(1,0)^2+\Gamma(0,1)^2-\Gamma(2,0)-\Gamma(0,2)\Big)+\frac{1}{2\tau}\Big(A_\Gamma\cos 2\theta+B_\Gamma\sin 2\theta\Big),
        \label{hstar_2}
\end{split} 
\end{equation}
where
\begin{equation}
    A_\Gamma=\frac{1}{2}\Big(\Gamma(1,0)^2-\Gamma(0,1)^2-\Gamma(2,0)+\Gamma(0,2)\Big), \quad B_\Gamma=\Gamma(1,0)\Gamma(0,1)-\Gamma(1,1)+\frac{i}{2}.
\end{equation}
Note that both $A_\Gamma$ and $B_\Gamma$ are real. We define $A_\Gamma = R_\Gamma\cos\phi_\Gamma$ and $B_\Gamma = R_\Gamma\sin\phi_\Gamma$ such that $R_\Gamma=\sqrt{A_\Gamma^2+B_\Gamma^2}\geq0$. Then Eq.~\eqref{hstar_2} can be written as
\begin{equation}
  \begin{split}\mathcal{H}^\star=&\lambda_1^{\max}|\kappa(2,0)|+\lambda_2^{\max}|\kappa(3,0)|-\frac{1}{2}\Big(\kappa(2,0)+\kappa(0,2)\Big)+\frac{1}{2\tau}\Big(\Gamma(1,0)^2+\Gamma(0,1)^2-\Gamma(2,0)-\Gamma(0,2)\Big)\\&+\frac{1}{2\tau}R_\Gamma\cos(2\theta-\phi_\Gamma).
        \end{split}\label{hstar_3}
\end{equation}
Assuming $\phi_\Gamma\in [-\pi,\pi]$ and $\theta\in[-\tfrac{\pi}{2},\tfrac{\pi}{2}]$, the supremum for the above Hamiltonian is achieved for $\theta^\star=\frac{1}{2}\phi_\Gamma$,
 where the value of $\phi_\Gamma$ is determined by $A_\Gamma$, $B_\Gamma$ and their signs. The supremum value of the Pontryagin Hamiltonian is

 \begin{equation}
 \begin{split}
     \mathcal{K}(\hat{\sigma},\hat{\rho})&=\lambda_1^{\max}|\kappa(2,0)|+\lambda_2^{\max}|\kappa(3,0)|+\frac{1}{2\tau}R_\Gamma-\frac{1}{2}\Big(\kappa(2,0)+\kappa(0,2)\Big)\\&+\frac{1}{2\tau}\Big(\Gamma(1,0)^2+\Gamma(0,1)^2-\Gamma(2,0)-\Gamma(0,2)\Big).
        \label{hstar_3_app}
 \end{split}\end{equation}
 From Eqs.~\eqref{dGamnmdt} and \eqref{dkapnmdt} we can calculate the evolution of the coefficients appearing above.  Then, 
 \begin{subequations}
     \begin{align}
         \begin{split}
             \frac{d\Gamma(1,0)}{dt}=&\Gamma(0,1) -
             \frac{\sin\theta}{4\tau}\left(\cos\theta\kappa(1,0)+\sin\theta\kappa(0,1)\right),
         \end{split}\\
          \begin{split}
             \frac{d\Gamma(0,1)}{dt}=&-(1+2\lambda_1)\Gamma(1,0)- 3\lambda_2\Gamma(2,0)+
             \frac{\cos\theta}{4\tau}\left(\cos\theta\kappa(1,0)+\sin\theta\kappa(0,1)\right),
         \end{split}\\
         \begin{split}
             \frac{d\kappa(1,0)}{dt}=&\kappa(0,1),
       \quad 
             \frac{d\kappa(0,1)}{dt}=-(1+2\lambda_1)\kappa(1,0)-3\lambda_2\kappa(2,0).
         \end{split}
     \end{align}
 \end{subequations}
 We denote $\tilde{\Gamma}(1,1)=\Gamma(1,1)-\frac{i}{2}$, $\tilde{\Gamma}(2,1)=\Gamma(2,1)-i\Gamma(1,0)$, $\tilde{\kappa}(2,1)=\kappa(2,1)-i\kappa(1,0)$ to ensure the scalar variables are all real. The second order $\Gamma$ terms evolve as 
\begin{subequations}
     \begin{align}
         \begin{split}
             \frac{d\Gamma(2,0)}{dt}=&2\tilde{\Gamma}(1,1) +
             \frac{\sin\theta}{2\tau}\Bigg(\cos\theta\Big(\Gamma(1,0)\kappa(1,0)-\kappa(2,0)\Big)+\sin\theta\Big(\Gamma(0,1)\kappa(1,0)-\kappa(1,1)\Big)\Bigg),
         \end{split}\\
          \begin{split}
             \frac{d\tilde{\Gamma}(1,1)}{dt}=&-(1+2\lambda_1)\Gamma(2,0)+\Gamma(0,2)-3\lambda_2\Gamma(3,0) \\&
             +\frac{1}{4\tau}\Bigg(r_\theta\Big(\sin\theta\kappa(0,1)-\cos\theta\kappa(1,0)\Big)+\Big(\cos^2\theta\kappa(2,0)-\sin^2\theta\kappa(0,2)\Big)\Bigg),
         \end{split}\\ \begin{split}
             \frac{d\Gamma(0,2)}{dt}=&-2(1+2\lambda_1)\tilde{\Gamma}(1,1)- 6\lambda_2\tilde{\Gamma}(2,1)
             \\&+\frac{\cos\theta}{2\tau}\Bigg(\Big(\sin\theta\kappa(0,2)+\cos\theta\kappa(1,1)\Big)-\Big(\cos\theta\Gamma(1,0)+\sin\theta\Gamma(0,1)\Big)\kappa(0,1)\Bigg),
         \end{split}
          \end{align}
 \end{subequations}
 The second-order $\kappa$ terms evolve according to the following equations.
 \begin{subequations}
 \begin{align}
           \begin{split}
             \frac{d\kappa(2,0)}{dt}=&2\kappa(1,1) +
             \frac{2\sin\theta}{\tau}\Bigg(\cos\theta\Gamma(2,0)+\sin\theta\tilde{\Gamma}(1,1)-r_\theta\Gamma(1,0)\Bigg),
         \end{split}\\
          \begin{split}
             \frac{d\kappa(1,1)}{dt}=&-(1+2\lambda_1)\kappa(2,0)+\kappa(0,2)-3\lambda_2\kappa(3,0)\\& +
             \frac{1}{\tau}\Bigg(r_\theta\Big(\cos\theta\Gamma(1,0)-\sin\theta\Gamma(0,1)\Big)+\Big(-\cos^2\theta\Gamma(2,0)+\sin^2\theta\Gamma(0,2)\Big)\Bigg),
         \end{split}\\ \begin{split}
             \frac{d\kappa(0,2)}{dt}=&-2(1+2\lambda_1)\kappa(1,1)- 6\lambda_2\tilde{\kappa}(2,1)\\&
             +\frac{2\cos\theta}{\tau}\Bigg(-\Big(\sin\theta\Gamma(0,2)+\cos\theta\tilde{\Gamma}(1,1)\Big)+\Big(\cos\theta\Gamma(1,0)+\sin\theta\Gamma(0,1)\Big)\Gamma(0,1)\Bigg).
         \end{split}
     \end{align}
 \end{subequations}
 
\section{\label{numerics_OC}Numerical Methods for finding optimal control}We present the numerical methods adopted for finding the optimal $\theta(t)$ and $\lambda_1(t)$ for the monitored oscillators presented in Sec.~\ref{control_SHO} (e.g.~Figure~\ref{fig:optimal_control_binomial_code}). 
 To find the optimal controls, Eqs.~\eqref{first_G_k_txt},\eqref{secondorder_G_txt}, \eqref{secondorder_k_txt}  are integrated with the optimality conditions presented in Eqs.~\eqref{l1_opt_txt} and \eqref{theta_opt_txt}  taken into account.  We assume initial values of the 10 scalars $\Gamma(1,0)$, $\Gamma(0,1)$, $\kappa(1,0)$, $\kappa(0,1)$, $\Gamma(2,0)$, $\tilde{\Gamma}(1,1)$, $\Gamma(0,2)$, $\kappa(2,0)$, $\kappa(1,1)$ and $\kappa(0,2)$. The density matrix update at each time step is calculated according to the steps presented in Appendix \ref{numerics_theta0} with a fourth-order Runge-Kutta method \cite{press2007numerical}.  We express the density matrix in the Fock state basis up to level $n=35.$ Since we consider fixed endpoint problems, solving for the optimal paths is the same as finding the initial values of $\Gamma$ and $\kappa$ such that at the final time $t_f$, the target state $\hat{\rho}_f$ is reached.  To that end, we maximize the fidelity in Eq.~\eqref{def_fidelity} between the target state $\hat{\rho}_f$ and the final state achieved $\hat{\rho}(t_f)$. The maximization is performed with respect to the initial values of the parameters $\Gamma$ and $\kappa$. For maximization, we adopted simulated annealing \cite{dowsland_experiments_1993} using  JAX \cite{jax2018github}. As mentioned previously, Pontryagin's maximum principle provides a necessary but not sufficient condition for optimality. Thus, integrating Eqs.~\eqref{first_G_k_txt}, \eqref{l1_opt_txt}, \eqref{theta_opt_txt},  \eqref{secondorder_G_txt}, \eqref{secondorder_k_txt}  does not guarantee that the obtained solutions are optimal. To ensure optimality (or near optimality), after a certain fidelity (Eq.~\eqref{def_fidelity}) is reached (90-95\%), we only accept solutions if the probability cost function in Eq.~\eqref{Jcost} is further reduced. Therefore, the problem becomes a Pareto front optimization between the fidelity with respect to the target state and readout probabilities. Robust numerical methods to find solutions that are guaranteed to be optimal should be investigated. The codes for finding optimal control are available at  \cite{Tathagata_CDJ-P_Monitored_Oscillator_2025}.
 
 \section{\label{numerics_trajectory}Numerical method for generating stochastic trajectories}We present the numerical methods adopted for generating stochastic trajectories with given initial state $\hat{\rho_i}$ and control.  We use the It\^{o}  form of the stochastic master Eq.~\eqref{SME_SHO}, given by
 \begin{equation}
\begin{split}
     \frac{\partial\hat{\rho}}{\partial t}
    =-i[\hat{H},\hat{\rho}] +\frac{1}{4\tau}\mathcal{D}_{\hat{L}_\theta}[\hat{\rho}]+\frac{1}{2\tau}\Big[\Delta\hat{L}_\theta,\hat{\rho}\Big]_+dW,
\end{split}
    \label{SME_SHO_ITO}
\end{equation}
with 
\begin{equation}
    \mathcal{D}_{\hat{L}_\theta}[\hat{\rho}] = \hat{L}_\theta\hat{\rho}\hat{L}_\theta-\frac{1}{2}\hat{L}_\theta^2\hat{\rho}-\frac{1}{2}\hat{\rho}\hat{L}_\theta^2.
\end{equation}
To preserve the positivity, we approximate the stochastic master equation as
\begin{equation}
    \hat{\rho}(t+dt)\approx \frac{\hat{\rho}^\prime}{\textrm{Tr}\hat{\rho}^\prime}, \quad \textrm{with} \quad \hat{\rho}^\prime=\hat{M}_{\textrm{It\^o}}\hat{\rho}\hat{M}_{\textrm{It\^o}}^\dagger, \quad \textrm{and} \quad \hat{M}_{\textrm{It\^o}} \approx \hat{\mathds{1}}-i\hat{H}dt+\frac{rdt}{2\tau}\hat{L}_\theta-\frac{dt}{8\tau}\hat{L}^2_\theta.
    \label{rho_prime_SME}
\end{equation}
 Eq.~\eqref{rho_prime_SME} is the same as \eqref{SME_SHO_ITO}  as the latter can be derived using the It\^o form Kraus operator  $\hat{M}_\textrm{It\^o}$  and expanding up to $\mathcal{O}(dt)$ with the It\^o rule $dW^2=dt$ in mind. At each time step, we generate a Gaussian random number with mean $0$ and variance $dt$ as the Wiener noise $dW$ and integrate Eq.~\eqref{rho_prime_SME}. 
\section{\label{numerics_sample_control}Numerical method for generating a sample control}In this section, we discuss the method of generating a sample control between $\hat{\rho_i}$ and $\hat{\rho_f}$ for a monitored harmonic oscillator. We first express the sample parameters as follows
\begin{equation}
    \theta(t)=\frac{\pi}{2}\tanh{\frac{2 f_1(t)}{\pi}}, \quad \lambda_1(t)=\lambda_1^{\max}\tanh{\frac{f_2(t)}{\lambda_1^{\max}}},
    \label{th_l1_f1_f2}
\end{equation}
where $f_1(t)$ and $f_2(t)$ are expressed in terms of Fourier series as below
\begin{equation}
f_1(t)=\sum_{n=0}^{N_\textrm{c}}\Bigg(c_n\cos\frac{2\pi n\Delta t}{T}+d_n\sin\frac{2\pi n\Delta t}{T}\Bigg), \quad f_2(t)=\sum_{n=0}^{N_\textrm{c}}\Bigg(c_n^\prime\cos\frac{2\pi n\Delta t}{T}+d_n^\prime\sin\frac{2\pi n\Delta t}{T}\Bigg),
\end{equation}
with $\Delta t =t-t_i$ and $T=t_f-t_i$. The hyperbolic tangents in Eq.~\eqref{th_l1_f1_f2} make sure the controls $\theta(t)$  and $\lambda_1(t)$ are bounded by $\frac{\pi}{2}$ and $\lambda_1^{\max}$. For $|\frac{2 f_1(t)}{\pi}| \ll 1$, $\theta_1(t)\simeq f_1(t)$. Similarly, for  $|\frac{f_2(t)}{\lambda_1^{\max}}| \ll 1$, $\lambda_1(t)\simeq f_2(t)$. For our simulations, $N_c=5$ is sufficient. We integrate Eq.~\eqref{op_rho_sho} (with the  Hamiltonian given by Eq.~\eqref{text_parametricH}) and Eqs.~\eqref{first_G_k_txt} for the controls in Eq.~\eqref{th_l1_f1_f2}. We maximize the fidelity defined in Eq.~\eqref{def_fidelity}  with respect to coefficients $\{c_n, d_n, c_n^\prime, d_n^\prime\}$ using simulated annealing. Assuming we can reach fidelity close to 100\%, the solution obtained this way is the most likely path between $\hat{\rho_i}$ and $\hat{\rho_f}$ under the controls 
 defined in Eqs.~\eqref{th_l1_f1_f2}. The controls obtained this way are usually not optimal, as shown in the text. 

\section{\label{xmeasurement}Position Measurement}Under position measurements only ($\theta(t) =0$ in Eq.~\eqref{l_theta}), as $t\to\infty$, the covariance matrix elements of a Gaussian state harmonic oscillator reach steady state values. Adopting the notation from Ref.~\cite{PRXQuantum.3.010327}, we denote $q_1 = \expval{\hat{X}}$, $q_2 = \expval{\hat{P}}$,  $q_3 = 2\, \textrm{Var} \hat{X}$, $q_4 = 2\, \textrm{Cov} \left(\hat{X},\hat{P}\right)$,  and $q_5 = 2\, \textrm{Var} \hat{P}$. The steady state  values of $q_3$, $q_4$ and $q_5$ are given by 
\begin{equation}
        \tilde{q}_3 =\sqrt{4\tau(\sqrt{1+4\tau^2}-2\tau)},\quad
        \tilde{q}_4 =\sqrt{1+4\tau^2}-2\tau,\quad
        \tilde{q}_5 =\sqrt{\frac{1}{\tau}(\sqrt{1+4\tau^2}-2\tau)(1+4\tau^2)}.
    \label{fpoints}
\end{equation}The final state is a squeezed state with squeezing parameter $\xi=Re^{i\Theta}$ such that
\begin{equation}
  \sinh 2R = \pm \sqrt{\left(
    \frac{\tilde{q}_5-\tilde{q}_3}{2}\right)^2+\tilde{q}_4^2},\quad
  \cosh 2R = \frac{\tilde{q}_5+\tilde{q}_3}{2},  
\label{snh2r}
\end{equation}
where $R = \tfrac{1}{2}\log\left(\sinh 2R+\cosh 2R\right)$.  We can write $
\xi = Re^{i\Theta} = \frac{R}{\sinh 2R}\left(\frac{\tilde{q}_5-\tilde{q}_3}{2}-i\tilde{q}_4\right)$. Note that either choice of sign in Eq.~\eqref{snh2r} leads to the same $\xi$. 

Now consider an optimal path starting from $\left(q_{1i},q_{2i}\right)$ and ending at $\left(q_{1f},q_{2f}\right)$ at $t=t_f$. The solution for the unique optimal path is given by \cite{PRXQuantum.3.010327}
\begin{equation}
    \begin{split}
        &q_1(t)=\Big(\frac{\alpha_1(\tilde{q}_3^2+\tilde{q}_4^2)}{8\tau} t+q_{1i}\Big)\cos{t}+\Big(\frac{\alpha_2(\tilde{q}_3^2+\tilde{q}_4^2)}{8\tau}t+q_{2i}+\frac{\alpha_1(\tilde{q}_3^2-\tilde{q}_4^2)}{8\tau}+\frac{\tilde{q}_3 \tilde{q}_4 \alpha_2}{4\tau}\Big)\sin{t},\\
        & q_2(t)=\Big(\frac{\alpha_2(\tilde{q}_3^2+\tilde{q}_4^2)}{8\tau}t+q_{2i}\Big)\cos{t}- \Big(\frac{\alpha_1(\tilde{q}_3^2+\tilde{q}_4^2)}{8\tau} t+q_{1i}+\frac{\alpha_2(\tilde{q}_3^2-\tilde{q}_4^2)}{8\tau}-\frac{\tilde{q}_3 \tilde{q}_4 \alpha_1}{4\tau}\Big)\sin{t},\\
    \end{split}
    \label{OP_sol}
\end{equation}
where, $\alpha_1$ and $\alpha_2$ are integration constants. Values of $\alpha$'s can be determined from the final conditions through the matrix equation
\begin{equation}
\begin{split}
    \mathcal{A}(t_f)\begin{pmatrix}\alpha_1\\ \alpha_2\end{pmatrix} = \begin{pmatrix}q_{1f}-q_{1i}\cos{t_f}-q_{2i}\sin{t_f}\\q_{2f}+q_{1i}\sin{t_f}-q_{2i}\cos{t_f}\end{pmatrix}.
   \end{split}
   \label{mat_eqn}
\end{equation}
\end{widetext}
The (always invertible) matrix $\mathcal{A}(\T_f)$ is given by
\begin{equation}
\begin{split}
    \mathcal{A}(t_f)=\frac{1}{8\tau}\begin{pmatrix} (\tilde{q}_3^2+\tilde{q}_4^2)t_f \cos{t_f} +(\tilde{q}_3^2-\tilde{q}_4^2)\sin{t_f} & ((\tilde{q}_3^2+\tilde{q}_4^2)t_f+2\tilde{q}_3 \tilde{q}_4)\sin{t_f}\\[8pt] (-(\tilde{q}_3^2+\tilde{q}_4^2)t_f+2\tilde{q}_3 \tilde{q}_4)\sin{t_f} & (\tilde{q}_3^2+\tilde{q}_4^2)t_f \cos{t_f} -(\tilde{q}_3^2-\tilde{q}_4^2)\sin{t_f} \end{pmatrix}.
    \end{split}
    \label{matrix}
\end{equation}
%and is always invertible.

\end{document}